\documentclass[12pt]{article}
\usepackage{latexsym}
\usepackage{amsmath}
\usepackage{epsf}
\usepackage{bm}
\usepackage{amsfonts}
\usepackage{amssymb}
\usepackage{graphicx,multicol}
\usepackage{psfrag}
\usepackage{color}
\definecolor{rosso}{cmyk}{0,1,1,0.4}
\definecolor{rossos}{cmyk}{0,1,1,0.55}
\definecolor{rossoc}{cmyk}{0,1,1,0.2}
\definecolor{blu}{cmyk}{1,1,0,0.3}
\definecolor{blus}{cmyk}{1,1,0,0.6}
\definecolor{bluc}{cmyk}{1,1,0,0.1}
\definecolor{verde}{cmyk}{0.92,0,0.59,0.25}
\definecolor{verdec}{cmyk}{0.92,0,0.59,0.15}
\definecolor{verdes}{cmyk}{0.92,0,0.59,0.7}

\input xy
\xyoption{all}

\def\circa#1{\,\raise.3ex\hbox{$#1$\kern-.75em\lower1ex\hbox{$\sim$}}\,}

\font\tenrsfs=rsfs10 at 12pt
\font\sevenrsfs=rsfs7
\font\fiversfs=rsfs5
\newfam\rsfsfam
\textfont\rsfsfam=\tenrsfs
\scriptfont\rsfsfam=\sevenrsfs
\scriptscriptfont\rsfsfam=\fiversfs
\def\mathscr#1{{\fam\rsfsfam\relax#1}}

\makeatletter

\newcommand{\be}{\begin{equation}}
\newcommand{\ee}{\end{equation}}
\newcommand{\beqy}{\begin{eqnarray}}
\newcommand{\eeqy}{\end{eqnarray}}
\newcommand{\p}{\partial}

\newcommand{\mx}{\mbox}
\newcommand{\mt}{\mathtt}

\newcommand{\al}{\alpha}
\newcommand{\bb}{\beta}
\newcommand{\ga}{\gamma}
\newcommand{\te}{\theta}
\newcommand{\vt}{\vartheta}

\newcommand{\de}{\delta}
\newcommand{\De}{\Delta}

\newcommand{\e}{\epsilon}

\newcommand{\Om}{\Omega}
\newcommand{\la}{\lambda}
\newcommand{\La}{\Lambda}

\newcommand{\ti}{\widetilde}
\newcommand{\bM}{\tilde{M}}

\newcommand{\ra}{\rightarrow}
\newcommand{\Ra}{\Rightarrow}
\newcommand{\im}{\Longleftrightarrow}
\newcommand{\rfi}{\right)}
\newcommand{\lf}{\left(}

\newcommand{\LF}{\left(}
\newcommand{\RF}{\right)}
\newcommand{\LT}{\left[}
\newcommand{\RT}{\right]}
\newcommand{\ie}{{\it i.e.\ }}

\newcommand{\vs}{\vspace{5mm}\\}

\begin{document}
\tolerance=100000
\thispagestyle{empty}
\setcounter{page}{0}

\vspace{1cm}

\begin{center}
{\LARGE \bf
 Nonlinear Structure Formation 
 \\ [0.15cm]
 and ``Apparent'' Acceleration:
 \\ [0.18cm]
an Investigation

}
\vskip 2cm

{
{\large Tirthabir Biswas\footnote{tirtho@hep.physics.mcgill.ca}}$^a$,
{\large Reza Mansouri\footnote{mansouri@ipm.ir}}$^{a,b}$,
 {\large Alessio Notari\footnote{notari@hep.physics.mcgill.ca}}$^{a}$,

\vskip 7mm
{\it $^a$ Physics Department, McGill University, \\ 3600 University Road, 
Montr\'eal, QC, H3A 2T8, Canada}
\vskip 3mm
{\it $^b$ 
Department of Physics, Sharif University of Technology, \\ Tehran, 11365-9161, Iran and Institute for Studies in Theoretical Physics and Mathematics (IPM), Tehran, Iran
} 
\vskip 3mm
}
\vspace{1cm}
{\large\bf Abstract}
\end{center}
\begin{quote}
{\noindent

We present an {\it analytically solvable} nonlinear  model of structure formation in a Universe with only dust. The model is an LTB solution (of General Relativity) and structures are shells of different density. We show that the luminosity distance-redshift relation has significant corrections at low redshift when the density contrast becomes nonlinear. A minimal effect is a correction in apparent magnitudes of order $\Delta m\simeq 0.15$. We discuss different possibilities that could further enhance  this effect and mimick Dark Energy.
}

\end{quote}

\newpage

\setcounter{page}{1}
\tableofcontents

%%%%%%%%%%%%%%%%%%%%%%%%%%%%%%%%%%%%%%%%%%%%%%%%%%%%%
\section{Introduction}

In recent years the exploration of the universe has 
provided detailed information about the distance-redshift relationship of many sources at very small redshifts as well as at redshifts of order unity.
These observations, {\it if} interpreted within the framework of a homogeneous Friedmann-Lema\^itre-Robertson-Walker (FLRW) metric indicate that the universe is presently undergoing a phase of accelerated expansion \cite{review}. Such an accelerated
expansion is most commonly interpreted as evidence  for the presence of a negative pressure 
component (``Dark Energy'') in the mass-energy density of the universe. This has given rise to the so-called ``concordance'' flat $\Lambda CDM$ model, where dark energy constitutes $70\%$ of the present energy budget in the form of a cosmological constant along with $30\%$ of matter. A second fact lending credence to the  $\Lambda CDM$ model is the observation that CMB excludes, with high precision, the presence of spatial curvature. Applying this to the late time universe in an FLRW model leads to a mismatch between the observed matter density and  the FLRW critical density. This mismatch can also be cured by the presence of a Dark Energy component.

However, from the point of view of fundamental physics  the presence of such a tiny cosmological constant is almost absurd: first, the scale is extremely low compared to any possible fundamental scale, and second it is tuned in such a way that it shows up just at $z\simeq 1$, where we live. This is so baffling\footnote{There have been some ideas/efforts to address these issues. For instance, see string-gas inspired  coupled quintessence models \cite{string} by one of the authors. However, it is fair to say that there is not yet a completely satisfactory way of solving both the ``smallness'' and the ``coincidence'' problems.} that before concluding that we are indeed facing this huge puzzle, it is worth making efforts to explore if the correct equations are being used to fit the data.

The homogeneous FLRW metric  certainly offers the simplest paradigm to interpret  cosmological observations. Indeed, the FLRW model is a very good approximation in the Early Universe as probed by the homogeneity of the CMB (density contrasts in  photons and dark matter are of the order  $10^{-5}$ and  $10^{-3}$ respectively).
However at low redshifts, the density contrast in matter grows up to values of ${\cal O}(1)$ and beyond: small scales become nonlinear first, and then more and more scales enter this regime. Today, scales of order of $60/h \, {\rm Mpc}$ (which is $2\%$ of the size of the horizon) have an average density contrast of order 1. Structures observed with density contrast larger than $0.1$ (so, already in a mildly nonlinear regime) extend to few hundreds of Mpc (around $10\%$ of the size of the horizon). It is thus not clear or obvious  that one can keep using  FLRW metric to interpret the large amount of  high precision experimental data collected in recent times in a regime where most of the mass in the Universe is clumped into structures or it is forming structures. 

Three are three main physical effects that we could be missing while using the FLRW model:
\begin{description}
\item[I.\ \ ] The ``overall'' dynamics of a Universe with these inhomogeneities could be significantly different from the FLRW dynamics. 
\item[II.\ ] Even if the dynamics is approximatively the same,  one has to wonder about the light propagation in a clumpy Universe: since all our conclusions about the dynamics are based on observations of light, this is of crucial importance.
\item[III.] The fact that we have just one observer could influence significantly the observations: we could live in an especially underdense or overdense region, which may induce large corrections on the observations.
\end{description}
In recent times there has been an on-going debate as to whether some of these effects may be large enough to give rise to an ``apparent acceleration'' leading us to believe in dark energy when there is none. The aim of this paper is to construct an {\it exact model of non-linear structure formation} where we can systematically study these effects. We wanted a model where one can study (possibly also {\it analytically}) the effects that inhomogeneities can cause on quantities like the {\it redshift} and the {\it angular distance}, thereby hopefully providing us with valuable insights to resolve the issue. The full problem of solving Einstein equations and light propagation 
for a realistic mass distribution is beyond present human 
possibilities, 
so some simplification is needed.
Our simplification is to assume spherical symmetry and in particular our model is based on Lema\^itre-Tolman-Bondi (LTB) metrics which are exact spherically symmetric solutions of general relativity with only dust. This approach has the obvious advantage of being exact: it does not use perturbation theory (as most of previous literature did to address {\bf I}), it allows to study light propagation without any FLRW assumptions (as opposed to literature addressing point {\bf II}), studying the corrections on the distance and also on the redshift of light (which has been ignored by most previous literature). The challenge however,   is to make these kind of models as realistic as possible, or at least be able to extrapolate from them the information that is relevant for our Universe.

So far such models have been used mostly to understand what happens to observations if we are living in a special position, a void for example~\cite{celerier,pm84,kl92,Tomita00,Tomita01,wiltshire05,moffat,alnes,abtt06,m05,ps06}, and therefore can at best be thought of as describing the local universe. It is interesting to go beyond this approach. Firstly, we may be missing some important effects (such as light propagating through several structures) by looking just at the local universe. Secondly, although these models show interesting effects capable of mimicking dark energy, it is not clear how realistic are these setups~\cite{bolejko,garfinkle} (also, some of these results have been questioned by~\cite{Flanagan}). On the other hand, we note that a different approach (aimed at solving question {\bf I}), based on back reaction of inhomogeneities on the overall dynamics of the Universe, suggests a much smaller correction at the first nontrivial order~\cite{Huiseljak,rasanen,KMNR,frysiegel}, although even in this framework some authors have argued in favour of a much larger effect~\cite{rasanen,Notari,KMR}. Thus it is imperative that we understand the origin of the large effects and in the process hopefully also bridge the gap between the perturbative back-reaction effects and the exact toy model approach.

It is clear that to address these issues we have to go beyond a ``local'' or a ``perturbative'' description. Thus, although we  work with the simplifying assumption of spherical symmetry,  we try to make our model realistic in the following 
sense. The universe  looks like an onion (and we frequently refer to it as the ``Onion model''): there is a homogeneous 
background density and on top of it density fluctuations as a function 
of the radial coordinate. Our model preserves large-scale homogeneity and  the shells are also distributed in a homogeneous way, making the picture a lot closer to the real world. One may wonder, nonetheless, that the centre is still a special point. In fact, we find that the centre has some special properties. For instance typically, the curvature tends to become large at the centre. In this case if we find significant corrections to, say, the redshift-luminosity distance relation near the centre, it is hard to distinguish whether the correction originates from the non-linear structure formation, or the ``excess curvature''. This therefore can jeopardise what we are trying to avoid, that of violating the cosmological principle. We deal with this problem by considering an observer who sits in a generic position and 
looks at  sources in the radial direction, along which we have periodic inhomogeneities. To our knowledge, the luminosity-redshift relation in LTB 
models has always been analyzed putting the observer at the centre of 
the  coordinates\footnote{Except for an analysis at small $z$ in \cite{humphreys}.}. For the first time we derive an exact expression for the luminosity distance of an object as seen by an off-centre observer\footnote{In this respect we note that~\cite{Flanagan} has shown that locally around the centre in the LTB model the deceleration parameter cannot be negative.  However, this result is only local  and so the universe may accelerate slightly far away from the observer.}. 

At early times (at last scattering for example) we assume the density 
fluctuations to have an amplitude of the order of the fluctuations 
observed in the CMB. Then we are able to follow exactly the evolution of 
density fluctuations. At redshifts of $z \gtrsim {\cal O}(10)$ the density contrast grows exactly as in a perturbed FLRW.
Then, when the density contrast becomes of  ${\cal O}(1)$ nonlinear clustering 
appears. So at low redshift the density does what is expected: 
overdense regions start contracting and they become thin shells (mimicking structures), while underdense 
regions become larger (mimicking voids) and eventually they occupy most of the 
volume. One can see this completely analytically upto high density contrasts (we have checked that the analytical solution is good even with $\de \rho/\rho\sim {\cal O}(100)$), which already makes it  an interesting model for the theory of structure formation. However, the really nice feature of the Onion model is that one can even calculate how the redshift and the various distances get corrected when we are in this non-linear regime.

In this model we can solve for the radial geodesics, therefore determining exactly what happens to the redshift of light. We
 place the observer in a generic location and measure the redshift as a function of the source coordinate. Then we trace a beam of light that starts from the source with some initial angle, and in particular we are able to compute the area seen by the observer. This provides us with the angular distance ($D_A$) or equivalently the luminosity distance ($D_L$) for a generic source.
In this way we provide the $D_L-z$ relationship (or the $D_A-z$ relationship) exactly. Thus we are able to study the three possible effects ({\bf I},{\bf II}, and {\bf III}) both numerically and  analytically. We now discuss our findings briefly.

The first issue ({\bf I}) of whether there is an overall effect in expansion due to inhomogeneities has been investigated  only recently. Buchert~\cite{buchert} has defined an average volume expansion and has developed a formalism to take into account of the backreaction of the inhomogeneities, although the problem of computing the size and the effect of the corrections is left unanswered. Rasanen~\cite{rasanen} has tried to estimate the effect of inhomogeneities (extending the formalism developed in~\cite{brand}, albeit in a different context) using perturbation theory, finding an effect of order $10^{-5}$ (in agreement with~\cite{Huiseljak}) and speculating about nonlinear effects. A complete second order calculation has been done in~\cite{KMNR}, and in a subsequent work one of us has shown~\cite{Notari} that the perturbative series is likely to diverge at small $z$ due to the fact that all the perturbative corrections become of the same size (each one is $10^{-5}$), and one may hope that these corrections give rise to acceleration (which it is shown to be in principle possible~\cite{nambu,Notari}). 
The fact that this may lead to acceleration has been explored in subsequent papers~\cite{KMR} and criticized in others (\cite{frysiegel} for example have argued the effect to be too small). Unfortunately a perturbative approach seem inadequate to answer this question as the phenomenon is intrinsically nonlinear.

On the other hand, since our model is an exact solution of  Einstein equations,  we can expect to get faithful results. Based on our analysis we do not find any significant overall effect\footnote{In agreement with the similar findings of~\cite{bolejko,Sugiura}}. Quantities, such as the  matter density, on an average, still behave as in the homogeneous Einstein-de Sitter (EdS) universe (\ie a matter dominated flat Universe). We remind the reader that if the universe starts to accelerate the matter density should dilute faster than EdS models. 

The second issue ({\bf II}) has been investigated in the past already in the '60s. The main point here is that the light that travels to us in a realistic Universe is not redshifted in the same way as in a FLRW metric, nor the angles and luminosities evolve in the same manner: it is more likely that a photon travels through almost empty space rather than meeting structures which are clumped in small sizes. However, typically work in the past has focused on what happens to the angles (or equivalently on luminosities), almost ignoring the problem of the redshift. Early pioneering work on this subject was due to Zeldovich~\cite{zeldovichL} that  estimated the effect on the angular distance ($D_A$) on light travelling through an empty cone embedded in an FLRW metric. 
A similar  treatment based on~\cite{dyerroeder} has been used recently to analyze data by the Supernova Cosmology Project Collaboration (see~\cite{Perlmutter}, fig.8). While giving a quite relevant correction (for large matter fraction $\Omega_m$), the effect does not allow to get rid of the cosmological constant.
Therefore the collaboration~\cite{Perlmutter} estimates to have control on the inhomogeneities. However such calculations are incomplete: the dynamics is completely ignored, and more importantly the redshift experienced by a light ray is assumed to be the usual FLRW result. 

Instead in the Onion model we could keep track of the corrections to both the redshift and the luminosity distance and in fact we found both of them to be significant, once the inhomogeneities become large. 
We found that the correction in the redshift with respect to a homogeneous model is of the same order, and even larger, than  the correction in luminosity distance.
Qualitatively one can understand where such corrections come from:  in the underdense regions the expansion is faster and the photons suffer a larger redshift than it would in the corresponding EdS model, while in the overdensities the expansion is balanced by the gravitational collapse and the photons experience milder redshift (or even a blue shift!). 
Here comes however one of the shortcomings of the Onion model: a radial light ray in the Onion model unavoidably meets underdense and overdense structures and they tend to average out the corrections.  In the real Universe, instead, where the light hardly encounters  any structure, one does not expect such cancellations to occur.
Nonetheless, based on these considerations, we can try to go beyond the Onion model: 
since the photon in the real world is mostly passing through voids it gets redshifted  faster as the non-linearities increase with time, thereby potentially producing the effect of apparent acceleration. In this paper we only present an estimate of such an effect leaving a more detailed study for  future work.

Next we come to the third issue ({\bf III}) of whether the position of the observer can have any important consequences. In general, in our analysis, we found that the corrections to redshift or luminosity distance is controlled by two quantities (in fact it is the product of the two): the amplitude of density fluctuations, $\de\rho/\rho$ and the ratio $L/r_{\mt{hor}}$, where $L$ is the wavelength of density fluctuations and $r_{\mt{hor}}$ is the horizon distance. As we mentioned earlier, the length scale at which the density fluctuations become non-linear, \ie $\de\rho/\rho\sim {\cal O}(1)$, the latter ratio is only around a few percent. Therefore naively one may expect these corrections to be too  small to be relevant for supernova cosmology (as argued in several literature). However the issue is more subtle: near the observer when the redshift and distances are themselves small these corrections become significant as they do not proportionately decrease. Instead of the ratio $L/r_{\mt{hor}}$, the relative correction is now governed by the ratio $L/\de r$, where $\de r$ is just the distance between the source and the observer. Thus in the first few oscillations, especially the first one, when $\de r\sim L$ we obtain significant corrections. 

In our analysis we find that, if we want to explain the Hubble diagram solely by the effect coming from the first  oscillation, then  the location of the observer is of crucial importance\footnote{This situation can change drastically if we are also able to account for the second effect, that of light mostly propagating through voids.}. Basically, what one needs to explain is the mismatch in the measurement of the local Hubble parameter (between redshifts $0.03<z<0.07$) and the luminosity distance observed for high redshift supernovae (typically between redshifts $0.4<z<1.5$). Although we can place the observer in any generic location, we find that the best choice, to account for the larger locally measured Hubble parameter, is if we  place the observer in an underdense region. Our main result is that we found that we require $3-4$ times larger local density contrast than the average value inferred by CMB measurements. 
 With these assumptions, however, we find that the Onion model can reasonably fit the supernova data and in this sense  our model  corroborates some of the findings of  the  models based on local voids.  We further find that the Onion model can be consistent with other observations such as local density measurements, the first peak position of the CMB (which measures the curvature of the universe) and baryon acoustic oscillations.  

The Onion model also offers us the opportunity to look at other relevant observations. For instance, since we have the freedom to choose the position of the observer we can in principle calculate quantities such as anisotropies in the measurement of Hubble constant which are essential to make a realistic assessment of the viability of any  model relying on  ``us'' being in a special position.  

Finally, suppose that inhomogeneities do not give a large enough effect to mimick Dark Energy, we can still estimate what is a minimal effect due to structures that typically affects distance measurements in the real Universe, \ie an universe constructed by just propagating the CMB fluctuations  to today.
We find that a typical effect gives a correction of order $0.15$ apparent magnitudes to the Hubble diagram at all redshifts. We can also explain another source of uncertainty in supernovae: the observed nearby supernovae seem to have larger intrinsic scatter~\cite{astier} which can easily be explained by the large oscillations that we get in the $D_L(z)$ curve close to the observer.  We briefly discuss such applications in this paper but mostly focus on the problem of dark energy. 

The structure of the paper is as follows.
In section~\ref{model} we introduce our LTB model, and in section~\ref{structure} we solve it and discuss structure formation.
In section~\ref{photon} we solve for light propagation in this background, both analytically and numerically. In section~\ref{luminosity} we discuss how to find the luminosity distance, both analytically and numerically.
In section~\ref{apparent} we  first construct a setup that could mimick Dark Energy, and then ask how realistic is such a setup, discussing observations such as local measurements of matter densities, CMB acoustic peaks and baryon oscillations. 
 We then go on to  discuss light propagation in the real Universe as opposed to the Onion model, and we estimate how large can the missing effect be. We also estimate  minimal uncertainties that inhomogeneities can lead to in the $D_L-z$ relationship.
Finally in section~\ref{conclusions} we draw our conclusions.
Appendices contain many technical details: Appendix~\ref{Esmall} contains a discussion of the analytical approximations that we employ, Appendix~\ref{nearcentre} contains a brief discussion of what happens near the centre of the LTB model,
Appendix~\ref{tr} and~\ref{app-redshift} contain the analytical derivation of (respectively) the time and the redshift along the photon trajectory and Appendix \ref{OA} contains the full calculation of $D_L$ for an off-centre observer. In appendix \ref{sachs} we show how our results for the off-centre angular distance is consistent with the optical scalar evolution equations for the angular distance derived by Sachs in 1961~\cite{sachs}.

%%%%%%%%%%%%%%%%%%%%%%%%%%%%%%%%%%%%%%%%%%%
\section{The model: LTB Universe}
\label{model}
As we mentioned before, our model is based on an LTB metric. LTB metrics are spherically symmetric solutions of  Einstein's equations with only 
dust (pressureless matter) acting as the source of energy density. So, in this respect, they are similar to the EdS model, but one is allowed to have an inhomogeneous distribution of matter along the radial direction, and this is what we are going to exploit in order to describe structure formation. Such a solution was first proposed by Lema\^itre 
and it was later  discussed by Tolman and Bondi~\cite{LTB}. Early applications of LTB~\cite{LTB} models were concentrated 
on the study of voids~\cite{os79,osv83}.
% Later, others have studied the formation of non-linear density contrasts in a Newtonian approximation to a LTB model embedded in an otherwise homogeneous universe~\cite{os79} and the formation of voids in a general relativistic context~\cite{osv83}.
Raine and Thomas used a LTB model to 
study the CMB anisotropies produced by an overdense matter distribution having a small density contrast of about 1.5 percent on a scale of order 1000 Mpc~\cite{rth81}. This work has then been generalized to the non-linear cases in~\cite{afms93}. The problem of light 
propagation in an LTB model and its impact on the Hubble and deceleration parameter 
was dealt with in~\cite{pm84,EllisRev}.

After the release of the new data on SNIa, following Celerier's~\cite{celerier} interesting paper exploring the possibility that large scale inhomogeneities may be able to mimick dark energy, several authors have  used  LTB solutions to understand such  effects~\cite{Tomita00,Tomita01,wiltshire05,moffat,alnes,abtt06,m05,ps06,cgh05,Flanagan}.  Some authors have  tried to study the effect of averaging the Einstein equations for LTB inhomogeneous solutions and its resulting backreaction which may also mimick an accelerating universe~\cite{r04,m05,ps06,kknny06}, while~\cite{bolejko} has questioned such approaches, as requiring unrealistic density fluctuations. Our Onion  model however differs from these ``void models'' in the following sense. Firstly,  our  LTB solution incorporates the entire Universe, without giving any physical meaning to the existence of a centre. Secondly, the philosophy behind our model is very different as compared to the Void models. We start with an initial universe which resembles the CMB epoch and then evolve to our present era, while the void models are typically valid only at late times.
 Also, very importantly we are able to analytically understand all our results thereby clarifying several issues that have been debated in the literature.

There are, however, limitations to our model of structure formation.
The most apparent limitation is that we can describe only 2-dimensional 
structures (that look like thin spherical shells).
While this is not completely realistic, it is nonetheless similar to 
what happens in the study of structure formation: using for example the 
so-called Zeldovich approximation~\cite{zeldovich} it can be shown that 
structures form after the collapse along one of the dimensions, leading to 
two-dimensional pancake-like objects.

The really crucial limitation of our model instead is that, as in the 
Zeldovich approximation once these structures start collapsing, there 
is nothing that can prevent them to become infinitely thin (shell 
crossing).
This is a serious limitation since, in the real world, objects which 
reach high density at some point can virialize and form stable 
structures.
The reason why our model does not include this feature is that we have 
motion only in the radial direction (and no pressure whatsoever) and 
therefore there is nothing that can prevent collapse: in the real world 
rotational motion is allowed, and systems can virialize if centrifugal 
forces balance the gravitational attraction.

In order to avoid this singular collapse, we choose a scale with an 
initial density contrast such that the collapse happens only in the 
future 
with respect to observation time.
This forces the redshift, $z_{nl}$, at which the density become 
nonlinear, to be quite close to zero. In other words we cannot track  non-linearities for too long time (upto around $z\sim 1$)

We will come back to discuss these limitations later in section~(\ref{conj}), but for now our aim is to construct an LTB model which can resemble a realistic universe with a largely homogeneous component and evolving density fluctuations on top of it. 
%%%%%%%%%%%%%%%%%%%%%%%%%%%%%%%
\setcounter{equation}{0}
\subsection{LTB metrics}
We start with a brief review of LTB solutions. 
The LTB metric (in units $c=1$) can be written as
\begin{equation}
ds^2=-dt^2 + S^2(r,t)dr^2 + R^2(r,t)(d\theta^2 + \sin^2 \theta 
d\varphi^2) \, , \label{eq:14}
\end{equation}
where we employ comoving coordinates ($r,\theta,\varphi$) and
proper time $t$. 

Einstein's equations with the stress-energy tensor of a 
dust, imply the following constraints:
\begin{eqnarray}
S^2(r,t) &=& {R^{'2}(r,t)\over {1+2E(r)}} , \label{eq:15}   \\
{1\over 2} \dot{R}^2(r,t) &-& {GM(r)\over R(r,t)}=E(r) \, , \label{eq:16} 
\\
4\pi \rho (r,t) &=& {M'(r) \over R'(r,t) R^2(r,t)} \, , \label{eq:17}
\end{eqnarray}
where a dot denotes partial differentiation with respect to $t$ and a prime 
with respect to $r$. $\rho (r,t)$ is the energy density of the matter, 
and $G\equiv 1/m_{Pl}^2$ is the Newton constant. 
The functions $E(r)$ and $M(r)$ are left arbitrary. The function $E(r)$ is  related to the spatial curvature, while 
$M(r)$ approximately corresponds to the mass inside a sphere of comoving 
radial coordinate $r$ \cite{celerier}.

One easily verifies that ~(\ref{eq:16}) has the following solutions 
for 
$R(r,t)$, which differ owing to the sign of the function $E(r)$:
\begin{itemize}
\item For $E(r)>0$, 
\begin{eqnarray}
R&=&{GM(r)\over 2E(r)} (\cosh u-1) , \label{eq:18} \\
t-t_b(r)&=&{GM(r)\over [2E(r)]^{3/2}}(\sinh u-u) \, , \nonumber \\
\nonumber
\end{eqnarray}
\item  $E(r)=0$, 
\begin{eqnarray}
R(r,t)=\left[ {9GM(r)\over 2}\right]^{1/3} [t-t_b(r)]^{\frac{2}{3}} \, , 
\label{eq:19}
\nonumber
\end{eqnarray}
\item and $E(r)<0$, 
\begin{eqnarray}
R&=&{GM(r)\over -2E(r)}(1-\cos u) ,  \label{eq:20} \\
t-t_b(r)&=&{GM(r)\over [-2E(r)]^{3/2}} (u-\sin u) \, , \nonumber \\
\nonumber
\end{eqnarray}
\end{itemize}
where $t_b(r)$, often known as the ``bang-time'', is another arbitrary 
function of $r$. It is 
interpreted, for 
cosmological purposes, as a Big-Bang singularity surface  at which $R(r,t)=0$. This is analogous to the scale-factor 
vanishing at the big bang singularity in the homogeneous FLRW models. One can choose $t_b(r)=0$ at 
the symmetry centre $(r=0)$ by an appropriate translation of the $t=$ 
constant surfaces and describe the universe by the $t>t_b(r)$ part of 
the $(r,t)$ plane, increasing $t$ corresponding to the future direction. 

To summarize,  LTB models contain three arbitrary functions, $M(r),\ E(r)$ and $t_b(r)$. It is easy and instructive to check how to recover the homogeneous 
limit.
First of all, one has to choose the big bang time to be the same 
everywhere: conventionally $t_b(r)=0$.
Then the $E(r)$ function has to be chosen as:

\be
 \left\{\begin{array}{rl}

E(r)&= -E_0 r^2                   \qquad\hbox{Closed FLRW} \\
E(r)&= \phantom{-}0\phantom{-E}\qquad  \hbox{Flat FLRW} \\  
E(r)&= \phantom{-}E_0 r^2                    \qquad\hbox{Open FLRW} 
\end{array}\right.
\label{curvature}
\ee
where $E_0$ is any positive number. One immediately recovers the homogeneous FLRW models, with $E_0$  parameterizing the amount of 
curvature in the model.
%%%%%%%%%%%%%%%%%%%%%%%%%%%%%%%%%%%%%%%%%%%%%%%%%
\subsection{Choice of $M(r)$, $E(r)$ and $t_b(r)$}

First of all, the reader should not be confused by the terminology 
which is usual in the literature: models with $E(r)<0,\ E(r)=0,\ E(r)>0$ 
are 
called respectively open, flat and closed LTB.
However as we are going to show, a model with $E(r)>0$ can mimick easily 
a flat FLRW model, and can be consistent with measurements of the first 
CMB peak that tell us that ``the Universe is flat''.

As a first remark we note that one of the three functions that describe 
an LTB model is just a ``gauge'' degree of freedom. In fact, by an 
appropriate rescaling of the radial coordinate $r$ the function $M(r)$ 
can  always be set to  any monotonic function.
For example, starting with a radial coordinate $\tilde{r}$ and a 
generic function $\ti{M}(\tilde{r})$, one can perform a coordinate 
transformation $r=r(\ti{r})$ without affecting the form of the metric, 
\ie
\be
M(r)=\ti{M}(\ti{r}),\ E(r)=\ti{E}(\ti{r})\mx{ and 
}R(r,t)=\ti{R}(\ti{r},t) \, .
\ee
Thus we are free to define a new coordinate as follows:
\be
r(\tilde{r})\equiv \left(\frac{3}{4 \pi} \frac{M(\tilde{r})}{M_0^4} 
\right)^{1/3}\Ra M(r)=\frac{4 \pi}{3}M_0^4r^3 \, ,
\ee
where $M_0$ is an arbitrary mass scale.
In the rest of the paper we always use this convention. Thus, we are 
left just with two functions: $E(r)$ and $t_b(r)$.

It is easy to see that the function $t_b(r)$ becomes negligible at late 
times: $t\gg t_b(r)$. 
Therefore the function $t_b(r)$ can only represent a decaying mode in 
the density fluctuation\footnote{It is easy to check that no matter what function we use as $t_b(r)$ the universe becomes more and more homogeneous with time (with trivial $E(r)$), which is exactly the opposite of what we want.}. Since the information contained in 
$t_b(r)$ 
gets lost at late times, we can just ignore this function and set $t_b(r)=0$ 
for all $r$.
Instead, in order to describe mass condensation and void formation at 
late times, we  have to choose a non-trivial $E(r)$.

At this point we have the choice between $E(r)>0$ or $E(r)<0$\footnote{Choosing $E(r)=0$ gives us back the flat homogeneous EdS model.}. It is 
also possible in principle to have a mixed model with some regions where 
$E(r)<0$ and other regions where $E(r)>0$. For simplicity,  we make the choice  
to  
always have $E(r)> 0$. As we will show, this can describe mass 
condensation and at the same time  be consistent with the CMB first 
peak 
measurements. 

For later convenience we enumerate  the  simplified equations for our model with $t_b(r)=0$ 
\be
R(r,t)=r{4\pi (\bM r)^2\over 6 E(r)}(\cosh u-1) \, ,
\ee
\be
t=r{4\pi (\bM r)^2\over 3[2 E(r)]^{3/2}}(\sinh u-u) \label{tu} \, ,
\ee
where we have defined
\be
\bM\equiv {M_0^2\over m_{Pl}} \, .
\ee

%%%%%%%%%%%%%%%%%%%%%%%%%%%%
\setcounter{equation}{0}
\section{Structure Formation}    \label{structure}
In this section we will try to understand whether open LTB metrics can 
mimick homogeneity over large distance scales, as well as describe 
inhomogeneities (representing galaxies, clusters of galaxies, etc..) 
appropriately. In other words  we will try to single out the $E(r)$'s which can ``approximate'' the density profile of our universe.
%%%%%%%%%%%%%%%%%%%%%%%%%%%%%%%%%%%%%%%%%%
\subsection{Large scale homogeneity and ``small $u$'' approximation}
Throughout the entire paper we will be employing what we call the ``small $u$'' approximation.  In appendix \ref{Esmall} we show how this is  a 
very good approximation for a space-time region large enough  to 
encompass our entire causal horizon. The strength of the approximation lies in the fact that it can accurately describe the dynamics even when $\de \rho/\rho \gg 1$, and thereby it lets us study the effect of non-linear structure formation on the different physical quantities. In fact, it is mainly the validity of this approximation along with an adiabatic approximation scheme (see appendix \ref{tr} for details) that provides us with so much analytic control.

In  ~(\ref{tu}), if we only keep next to leading terms in $u$ we have
\be
t\approx{4\pi \bM^2r^3\over 3[2 E(r)]^{3/2}}\left({u^3\over 
6}+{u^5\over 5!}\right)\im \left({9[2 E(r)]^{3/2}t\over 2\pi \bM^2 
r^3}\right)^{1/3}=u\left(1+{u^2\over 20}\right)^{1/3} \, ,
\label{t-u}
\ee
and 
\be
R(r,t)\approx{4\pi \bM^2r^3\over 6E(r)}\left({u^2\over 2}+{u^4\over 
4!}\right) \, .
\label{R-u}
\ee 
It is clear from~(\ref{t-u}) that this approximation is valid (\ie $u$ 
is small) when 
\be
v\equiv \left({9[2E(r)]^{3/2}t\over 2\pi \bM^2 
r^3}\right)^{1/3}={\sqrt{E}\over \bM r}\left({9\sqrt{2}t\bM\over \pi 
}\right)^{1/3}\ll1 \, .
\label{v}
\ee
Now, if we choose an $E(r)$ that grows less than quadratically then, as long as we are sufficiently far away from the centre, 
we can trust  the small $u$ approximation  (see appendix \ref{Esmall} for a more 
precise estimate). 
We will see later that, in fact, we do not want an $E(r)$ growing faster (or equal) than $r^2$ because it spoils the large-scale homogeneity or it introduces unwanted overall spatial curvature. 

Proceeding accordingly, (\ref{t-u})  yields
$$v\left(1-{u^2\over 60}\right)\approx u$$
which can be solved  giving us
\be
u\approx v-{v^3\over 60} \, .
\label{u-v}
\ee
Combining~(\ref{R-u}), (\ref{v}) and (\ref{u-v}) we find
\be
R(r,t)\approx (6\pi)^{1/3} r(\bM t)^{\frac{2}{3}} \left[1+\al{E(r)(\bM 
t)^{\frac{2}{3}}\over (r\bM)^2}\right] \, ,
\label{R}
\ee
where we have defined
\be
\al\equiv \left({81\over 4000\pi^2}\right)^{1/3} \, .
\ee
As one can easily check, the prefactor in the expression for $R(r,t)$ 
corresponds to the flat FLRW result while $E(r)$ 
corresponds to corrections in $R(r,t)$.

Next, let us try to understand how the density profile looks like. 
First  we will approximately compute the average density inside any 
sphere 
of radius $\bar{r}$. 
\be
\langle \rho \rangle ={M_{\mt{tot}}\over V_{\mt{tot}}}={\int dr 
R^2R'\rho(r,t)\over \int dr R^2R'}=M_0^4{\int dr r^2\over \int dR R^2}=M_0^4{\bar{r}^3\over R^3} \, ,
\ee
where we have assumed that $E(r)\ll1$ (see appendix~\ref{Esmall} for a 
justification).
The last equation can be written as:
\be
\langle \rho\rangle={M_0^4\over 6\pi(\bM t)^{2}}\left[1+\al{E(\bar{r})(\bM 
t)^{\frac{2}{3}}\over (\bar{r}\bM)^2}\right]^{-3} \, .
\label{density-avg}
\ee

Imposing large scale homogeneity basically means that  $\langle 
\rho\rangle$ should have a well defined limit as  $\bar{r}\ra \infty$. From 
the 
above expression it is clear that this happens if 
$$\lim_{\bar{r}\ra \infty}{E(\bar{r})\over (\bar{r}\bM)^2}\longrightarrow 
\mt{const.}$$
which means that $E(r)$ cannot  increase faster than $r^2$ as  $r\ra 
\infty$. 
Moreover, as we have pointed out in the previous section, a quadratic $E(r)$ 
corresponds to a model with constant spatial 
curvature~(\ref{curvature}). This is strongly constrained by CMB data 
\cite{WMAP}, because the angular diameter distance 
($D_A$) for an universe with spatial curvature at high redshift is very different from the flat case
 prediction, and the measurement of the first peak position strongly 
favours the 
flat model (or more precisely constrains the amount of constant spatial 
curvature to be very small). Therefore, in order to avoid curvature, $E(r)$ actually needs to increase slower than $r^2$ as  $r\ra \infty$. We will now further show that only a linearly growing $E(r)$ (times a factor that includes fluctuations)  guarantees a 
realistic   density profile (relative heights of peaks and troughs of the fluctuations) that does not 
change as we move outwards from the centre towards infinity. 
We will explicitly see in section \ref{1stpeak}, that such an $E(r)$ is also compatible with having a flat universe.

%%%%%%%%%%%%%%%%%%%%%%%%%%%%%%%%%
\subsection{Inhomogeneities}
To compute $\rho(r,t)$ we have to compute $R'$. From~(\ref{R}) we find
\be
R'(r,t)=(6\pi)^{1/3}(\bM t)^{\frac{2}{3}}\left[1+\al(\bM 
t)^{\frac{2}{3}}{\left(E(r)\over r\bM^2\right)'}\right] \, .
\ee
From~(\ref{eq:17}) we then find
\be
\rho(r,t)={M_0^4\over 6\pi(\bM t)^2\left[1+\al{E(r)(\bM 
t)^{\frac{2}{3}}\over (r\bM)^2}\right]^2\left[1+\al(\bM 
t)^{\frac{2}{3}}{\left(E(r)\over 
r\bM^2\right)'}\right]} \, .
\ee
We note that the correction term in the second factor of the 
denominator is actually proportional to $v^2$
$$
\al{E(r)(\bM t)^{\frac{2}{3}}\over (r\bM)^2}=\al\left({\pi\over 
9\sqrt{2}}\right)^{2/3}v^2\ll1 \, ,
$$
and therefore can be ignored consistently with our ``small $u$'' 
approximation.
Thus we have
\be
\rho(r,t)={M_0^4\over 6\pi(\bM t)^2\left[1+\al(\bM 
t)^{\frac{2}{3}}A(r)\right]}\,\, ,\mx{ where 
}A(r)\equiv\left({E(r)\over r\bM^2}\right)'
\label{matterdensity}  \, .
\ee

A few comments are in order. First, observe that the FLRW behaviour for 
the density is given by the prefactor multiplying the denominator. 
The fluctuations are provided by the presence of 
$A(r)$ in the denominator. It is also clear that if we do not want the 
relative peak to trough ratio to change, then the function $A(r)$ 
must 
be bounded
\be
A_{min}\leq A(r)\leq A_{max}   \, .
\ee
In other words there should not be any ``overall'' growth of $A(r)$ as $r$ increases. From the definition of $A(r)$ (\ref{matterdensity}) this means that $E(r)$ has to have an ``overall'' linear growth apart from the oscillatory fluctuations; if $E(r)$ grows faster or slower than linear, $A(r)$ starts to grow or shrink respectively.  

It is clear that the density contrast ($\de$) defined in the usual 
manner
\be
\de\equiv {\rho(r,t)-\langle \rho \rangle(t)\over \langle \rho 
\rangle(t)} \, ,
\ee
is controlled by
\be
\e(r,t)\equiv \al(\bM t)^{\frac{2}{3}}A(r)  \label{epsilont} \, .
\ee
One obtains a simple relation, ignoring the small correction of (\ref{density-avg})
\be
\de= -{\e\over 1+\e}=-{\al(\bM t)^{\frac{2}{3}}A(r)\over 1+\al(\bM 
t)^{\frac{2}{3}}A(r)}  \, .
\label{contrast}
\ee
Let us  assume, for the purpose of discussing the dynamics, that $A_{min}$ is negative. We will see in the next subsection that this will be required in order to avoid curvature. It is clear that the density fluctuates between
\be
\de_{min}(t)\equiv-{\al(\bM t)^{\frac{2}{3}}A_{max}\over 1+\al(\bM 
t)^{\frac{2}{3}}A_{max}}<\de<{\al(\bM t)^{\frac{2}{3}}(-A_{min})\over 
1+\al(\bM 
t)^{\frac{2}{3}}A_{min}}\equiv\de_{max}(t) \, .
\ee
When $A(r)$ is close to $A_{max}$ we have a void, while when it is 
close to  $A_{min}$, it signals an overdensity.
It is nice to see how the transition from linear to nonlinear regime 
occurs.
When $\epsilon\ll 1$, $\de\sim \e$ and we are in the linear regime: an initial density 
fluctuation grows at small times as 
$t^{\frac{2}{3}}$, in agreement with the prediction of cosmological 
perturbation theory.
When $\epsilon$ is no longer small the density contrast grows rapidly, 
and it is comforting to see that this result is  the same as found within 
the Zeldovich approximation \cite{zeldovich}.

Even if one chooses $A_{\mt min}=-A_{\mt max}$, in the nonlinear regime 
typically $|\de_{min}(t)|< \de_{max}(t)$. This is because the denominator in~(\ref{contrast}) comes in play. For positive $\e$ the denominator does not change much, but for negative $\e$ as it approaches $-1$, the denominator tends to zero thereby giving rise to a very large $\de_{max}$. In fact, as $t$ increases  there comes a point when the denominator in  ~(\ref{contrast}) 
precisely 
vanish so that the density in the over-dense region is infinite 
($\de_{max}(t)\ra \infty$), while $\de_{min}(t)$ remains finite. The 
``critical'' time when this happens is given by
\be
t_s={\bM^{-1}\over (-\al A_{min})^{3/2}}  \label{ts}  \, .
\ee
In the real Universe, rather than the density going to infinity, the 
system typically virializes, forms rotating structures (galaxies, 
clusters, etc.), and falls out of the Hubble expansion. As a result, 
$\de_{max}(t)$ in reality should saturate to a maximal value (unless a 
black hole 
is formed), while the underdense regions keep forming bigger and 
emptier voids. Unfortunately in our model we cannot incorporate the 
effects 
of virialization and cannot  trust the results beyond $t_s$. We will 
come back to the important question of whether this limitation may lead 
us 
to underestimate the effects of inhomogeneities on the 
luminosity-redshift relation in section~\ref{conj}.
%%%%%%%%%%%%%%%%%%%%%%%%%%%%%%%
\subsection{The ``Onion'' Model}  \label{onionmodel}
Let us come back to the main issue of finding  appropriate $E(r)$'s 
that can  describe the large scale structures. We have seen how the 
function $A(r)$ controls the density profile and in particular that we need a bounded $A(r)$. From now on we will focus on a model with sinusoidal density contrasts. (It is possible to consider a spectrum of many wavelengths, but in order 
to keep the analysis simple we will focus on a single wavelength $L$.)
 As we will justify later,  to achieve this it  is sufficient to obtain a density profile
that is sinusoidal in the comoving coordinate $r$. 

It is evident from~(\ref{epsilont}) that, to obtain such periodic 
oscillations of over and under dense regions, we require a periodic 
$A(r)$:
\be
A(r)=\left[\bar{A}_0+A_1\sin\LF{2\pi r\over L}\RF\right] \, .
\label{A}
\ee

Integrating~(\ref{A}) we find
\be
E(r)=\bM^2\left[r^2\bar{A}_0+rA_0-r{A_1L\over 2\pi}\cos\LF{2\pi r\over L}\RF\right] \, .
\ee
The term proportional to $\bar{A}_0$ gives rise to curvature and 
hence we do not consider it anymore. It is easy to see that this is precisely the term that one would generate if $A_{min}>0$ and this is why we said before that we needed $A_{min}<0$.  In the absence of $\bar{A}_0$, $E(r)$ becomes 
$$
E(r)=\bM^2\left[rA_0-r{A_1L\over 2\pi}\cos\LF{2\pi r\over L}\RF\right] \, .
$$
We note that the term proportional to $A_0$ does not enter into the 
density profile  and therefore we do not expect it to play any significant role in modifying the luminosity distance-redshift 
relation (we discuss this in sect.~\ref{photon2}). Since we want $E(r)$ 
to be always positive, it seems convenient to choose
\be
A_{0}={A_1 L\over 2\pi} \, ,
\ee
so that
\be
E(r)=\bM^2r{A_1L\over 2\pi}\left[1-\cos\LF{2\pi r\over L}\RF\right]=\bM^2r{A_1L\over 2\pi}\sin^2\LF{\pi r\over L}\RF
\label{cosineE} \, .
\ee
This defines our model uniquely, with  two parameters, $L$ and 
$A_1$. For later convenience we enumerate $R$ and $R'$ for our Onion model:
\be
R(r,t)\approx (6\pi)^{1/3} r(\bM t)^{\frac{2}{3}} \left[1+{\e\over 2\pi}{L\over  r}\sin^2\LF{\pi r\over L}\RF\right]
\label{onionR} \, ,
\ee
and 
\be
R'(r,t)=(6\pi)^{1/3}(\bM t)^{\frac{2}{3}}\left[1+\e \sin\LF{2\pi r\over L}\RF\right]
\label{onionRp} \, ,
\ee
where we have introduced the ``amplitude of oscillations'' 
\be
\e(t)\equiv \al A_1(\bM t)^{\frac{2}{3}}
\label{epsilon} \, .
\ee

It is useful to observe that at any given time-slice, $R$ is always proportional to the coordinate distance $r$, especially at earlier times when the  fluctuations due to inhomogeneities are small. Now, it is clear from the metric that $R$ is actually related to the proper distance; it is in fact exactly the space-like distance between two points along the same radial ray in a time slice. Thus to a good approximation, barring a proportionality factor, one can indeed treat the coordinate distance $r$ as the ``physical distance''. This explains why it physically makes sense to consider a density profile which is just periodic in $r$.

For an $E(r)$ given by~(\ref{cosineE}) the density profile looks like 
\be
\rho(r,t)={M_0^4\over 6\pi(\bM t)^2\left[1+\e(t)\sin\LF{2\pi r\over 
L}\RF\right]}\, , 
\ee
with periodic over-dense and void regions like an onion. The amplitude of density contrast is controlled by the function $\e(t)$ 
and increases as time progresses.

\begin{figure}
\hspace{-1.7cm}
 \psfrag{t5x1012sec}[][]{$t=5\times 10^{12}$ sec}
  \psfrag{t5x1016sec}[][]{$t=10^{16}$ sec} 
  \psfrag{t1017sec}[][]{$t=10^{17}$ sec} 
  \psfrag{t4}[][]{\, \, $\, \, t=4.7\times 10^{17}$ sec}
    \psfrag{r}[][]{$(r-r_O)/h$ Mpc}
  \psfrag{-}[][]{}
    \psfrag{rO}[][]{}
 \psfrag{delta}[][]{$\delta$}
 \psfrag{t}[][]{$t=5\times10^{12}$ sec}
\includegraphics[width=0.49\textwidth]{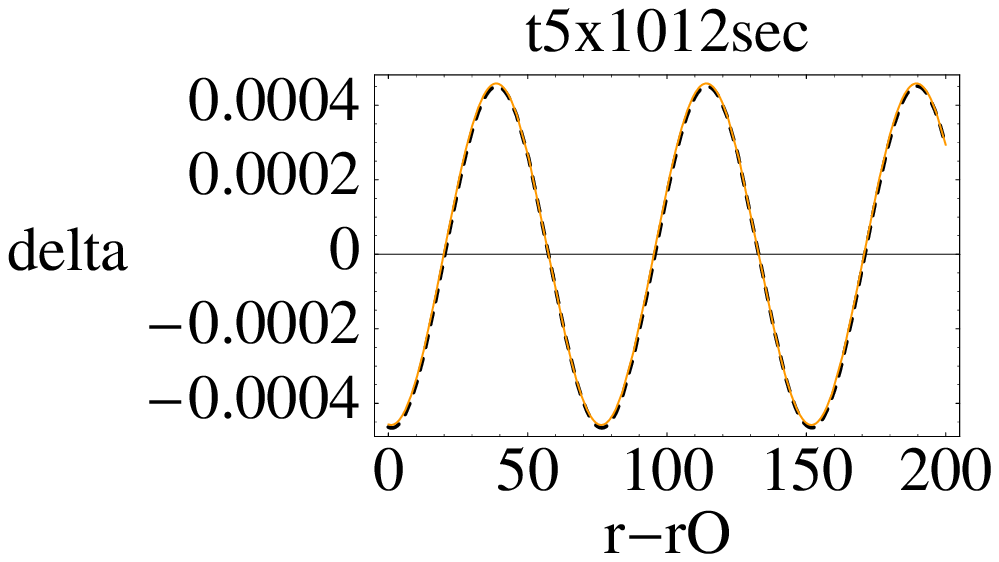}
\psfrag{delta}[][]{$\delta$}
 \psfrag{t}[][]{$t=5\times10^{16}$ sec}
 \hspace{-0.6cm}
\includegraphics[width=0.46\textwidth]{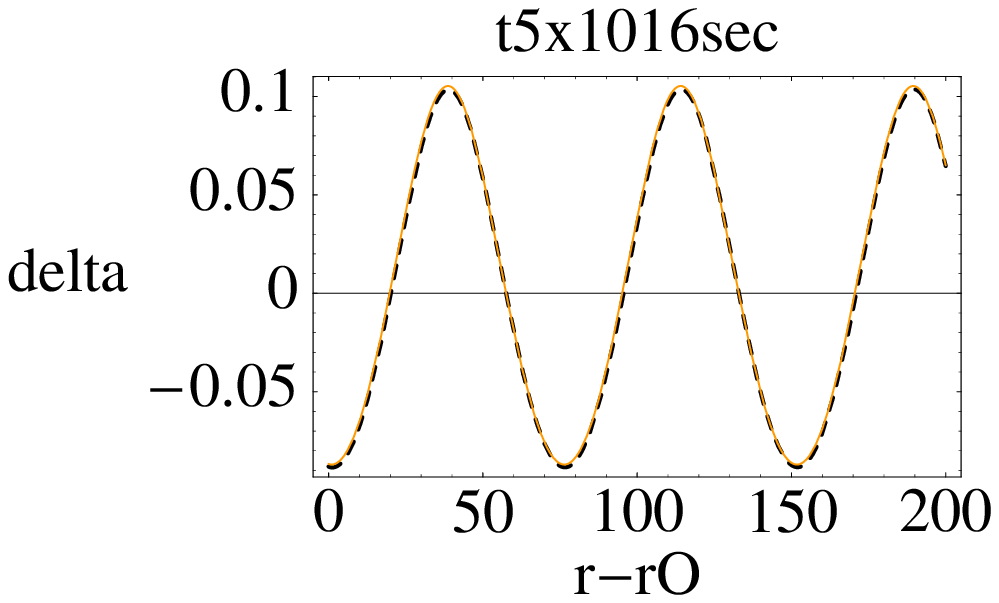}
\psfrag{delta}[][]{$\delta$}
 \psfrag{t}[][]{$t=5\times10^{17}$ sec}
\hspace{2cm}
\includegraphics[width=0.46\textwidth]{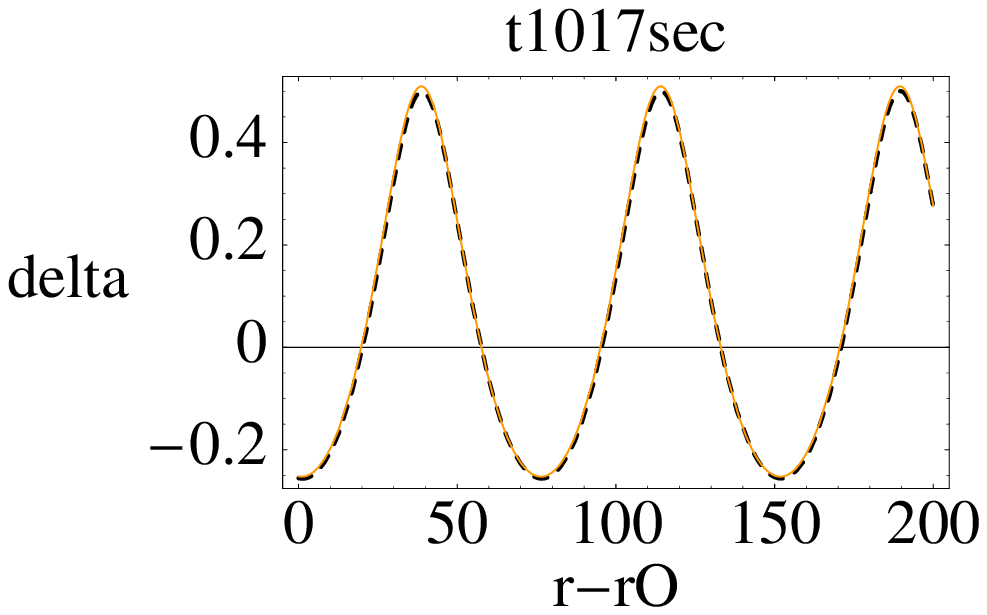}
\hspace{1.3cm}
\psfrag{delta}[][]{$\delta$}
 \psfrag{t}[][]{$t=5\times10^{17}$ sec}
\includegraphics[width=0.46\textwidth]{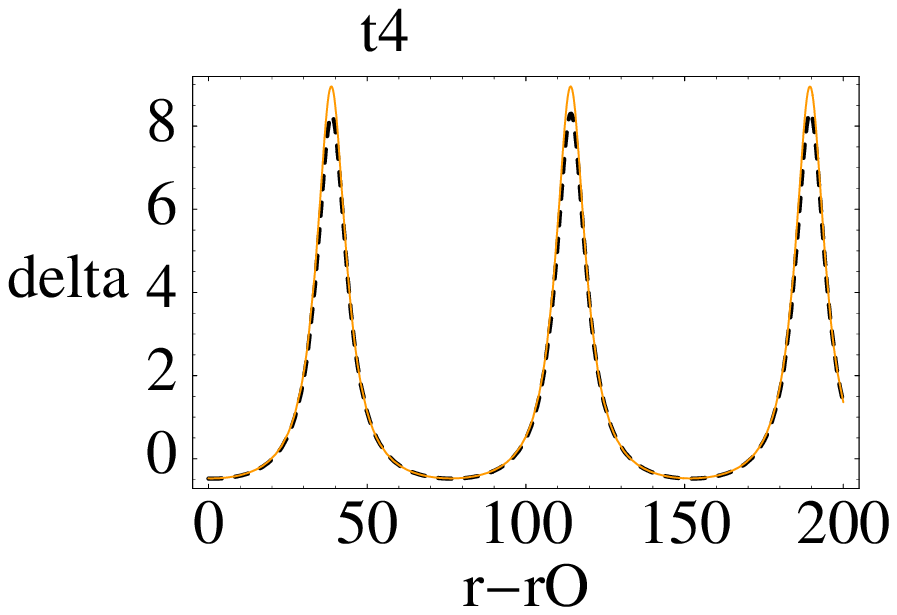}
\caption{\label{contrasto}
The density contrast for a given $L$ but at different epochs as a function of the radial 
coordinate, numerically (black dashed line) and analytically (orange 
solid line). The two curves are practically indistinguishable.
}

\end{figure}

As seen in figure~(\ref{contrasto}) our analytical estimate agrees with 
the exact numerical solution very well. As we discuss in details in  
appendix~\ref{Esmall}, this is possible because we can find a 
sufficiently large region in $r-t$ plane where both $E(r)$ and $u$ are 
small.

As anticipated, we encounter a problem of singularities, i.e. there is 
a time, $t_s$, at which the density contrast blows up. In our model
\be
A_{min}=-A_{max}=-A_1\mx{ with }A_1>0 \, ,
\ee
so the time $t_s$ is given by~(\ref{ts}):
\be
t_s\bM={1\over (\al A_1)^{3/2}} \, .
\ee
On the other hand we know the density contrast at CMB, as measured first 
by COBE\cite{COBE,WMAP}, is given by~(\ref{contrast})
\be
\de_{CMB}=\epsilon(t_{CMB})=\al A_1(\bM t_{CMB})^{\frac{2}{3}}
\label{A_1} \, .
\ee
We will abbreviate ``CMB'' by ``C'' from now on. Thus we find
\be
{t_s\over t_C}=\de_C^{-3/2}  \, ,
\label{t_s}
\ee
which means that the ``singularity'' time does not depend on the scale 
of the inhomogeneities ($L$), but only depends on the initial amplitude 
of density fluctuations. Clearly, this is also the time when the 
density contrast is just relevant for structure formation 
$\de\rho/\rho\sim 
{\cal O}(1)$, so we will choose $t_s$ to lie close to $t_O$, the observation time.

We stress again here the limitations of our model. Due to the fact that 
we work in the region $t<t_s$, we can have structure formation only at very low 
redshifts. In this sense the only scales that we can describe 
realistically are those that collapse roughly at the present epoch, 
that is with 
$L$ of the order of several Mpc$/h$ (where $h$ is related as usual to 
the measured value of the Hubble constant $H_0$ by $H_0=100\, h \, {\rm 
Km}\, {\rm sec}^{-1}{\rm Mpc}^{-1}$), that corresponds to the scale of 
the 
largest seen structures like voids or superclusters.
Moreover if we work at times $t<t_s$, where $\e(t)<1$, it is clear from 
 ~(\ref{contrast}), that the minimal density contrast that can be 
achieved in the Onion model is only $\delta=-0.5$.

%%%%%%%%%%%%%%%%%%%%%%%%%%%%%%%%%%%
\setcounter{equation}{0}
\section{Radial Photon Trajectories and Redshift} \label{photon}
In the previous section we have seen how our LTB Onion model 
can be used to describe the nonlinear large scale structure of our 
universe. Our aim now is to find corrections to the $D_L-z$ curve 
coming from inhomogeneities like voids and superclusters that we 
observe in our present universe.

In the Onion model structures are concentric shells of high density, and voids are
 replaced by the space between the shells. The physical setup that we consider is to have the observer
 far from the centre of the LTB patch in order to avoid as much as possible
any peculiar behaviour due to the fact that the centre is a special point.
Moreover in this way we can make the photon propagate in a region in which $u$ is always small, and 
%therefore we 
%can easily make the density profile regular (as discussed in section\ref{structure}) and at the same time find 
under good analytical control. 
So, the observer 
will be located at some nonzero $r_O$, and we will choose it to be large enough such that $u$ is small, but not too large so that $E(r)$  also remains small.
Then we will consider sources at variable position $r>r_O$, but 
always along a radial trajectory.
In this case we are able to easily solve the geodesic equations 
numerically and also to find analytic approximations. In the same 
configuration we will be able to define and compute the angular (and 
luminosity) 
distance in section~\ref{luminosity}. 

At this point one may wonder about other directions that an observer can look at. We do not consider them because approximately the picture looks the same and we do not expect any qualitatively new physics.  It is another matter that technically it is extremely hard to solve for such non-radial trajectories.

\subsection{Null Geodesic Equations} \label{photon2}
The first step towards obtaining a relationship between the luminosity 
distance and the redshift is to solve  the null geodesic equations. In 
other words, we have to obtain $t(r)$ for a photon trajectory 
converging at say $r=r_O$ at time $t=t_O$. 
(The subscript $O$ from now on will denote the observer).

The equation for this radial photon trajectory is simply given by
\be
{dt(r)\over dr}=-{R'(r,t(r))\over \sqrt{1+2E(r)}} \, .
\label{t-radial}
\ee 

 First, note that we recover the FLRW limit by letting $E(r)\ra 0$ here and in 
 ~(\ref{R}). 
In this case we find
\be
t(r)^{1/3}=t_O^{1/3}-\bb\bM^{2/3}(r-r_O)\equiv t_{F}^{1/3}(r) \label{tFLRW}  
\, ,
\ee
where $$\bb \equiv {(6\pi)^{1/3}\over 3} \, .$$
The nice feature of the Onion model is that it allows us to find an analytic approximation to~(\ref{t-radial}), in the small $u$ and small $E$ framework (again, we show in 
Appendix~\ref{Esmall} that in our region $E(r)$ is always small). The result, derived in Appendix~\ref{tr} is:
\be
t(r)^{1/3}=t_F^{1/3} \left[1+ \frac{\epsilon}{2 \pi}  \left( \frac{L}{r_{\rm 
hor}} \right) \cos\LF{\frac{2\pi r}{L}}\RF \right]
\label{tL(r)} \, ,
\ee
where we have assumed 
\be
\cos\LF{2\pi r_O\over L}\RF=0 \, ,
\ee
\ie the observer is located either at a maximum or a minimum of the density profile\footnote{The expression could be generalized for any observer location, but it would look more cumbersome and it would not add anything to the understanding of the physics.} and we have defined  $r_{\rm hor}(t)$ as the radial coordinate distance travelled by 
light at time $t$ in an FLRW Universe (which is basically the same as in our Onion model):
\be
 \bM r_{\rm hor}(t)\equiv \frac{(\bM t)^{1/3}}{\bb} \, .
 \label{horizon}
\ee 
 Let us pause for a moment to understand~(\ref{tL(r)}). Firstly, we point out that the radial dependence of $t(r)$ is most explicit in the expression~(\ref{x-approx}). The reason we have written it in the form~(\ref{tL(r)})  is because it physically clarifies on what quantities the correction really depends. For example, it is nice to see how the correction to the FLRW result in~(\ref{tL(r)}) is controlled  by the parameters of the model, $\e$ and $L$: $\epsilon$ is the amplitude of the fluctuations (that grows like $t^{2/3}$) and only when it becomes of ${\cal O}(1)$ one can see an effect. The effect also  increase as we look at higher  coherence scale, $L$,  although we should keep in mind that with increasing  wave-lengths $\e$ decreases and it is really the product which is important. The largest scales that are non-linear today are around $60/h$ Mpc, so that $L/r_{\mt{hor}}\sim 0.02$ and so is the product, giving us a correction of only a few $\%$. We note in passing that even for scales as high as $\sim 400/h$ Mpc the product remains roughly the same ($\e\sim 0.1$ and $L/r_{\mt{hor}}\sim 0.1$) and therefore are equally relevant. Later in the next subsection  we will see, how close to the observer, the ``relative correction'' gets enhanced from a few \% to almost 20\%. Finally,  to recover the radial dependence of the right hand side in~(\ref{tL(r)}) we point out that the quantities $\e$ and  $r_{\rm hor}$ depend on $t_F$ via~(\ref{epsilon}) and~(\ref{horizon}) and the radial dependence of $t_F$ is given in~(\ref{tFLRW}).

\begin{figure}
\psfrag{r}[][]{$(r-r_O)/h$ Mpc}
  \psfrag{-}[][]{}
    \psfrag{rO}[][]{}
\psfrag{t1017sec}[][]{$t/10^{17}$ sec}
    \psfrag{delta}[][]{$\sqrt{\langle\delta^2\rangle}=0.5$}
\includegraphics[width=0.78\textwidth]{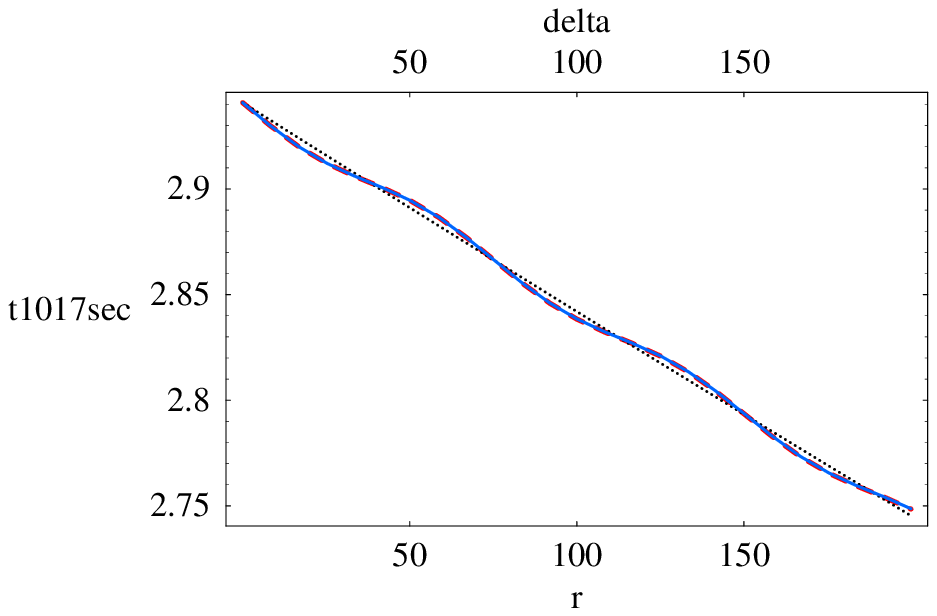}
\caption{\label{tempo}
{\small
Time along the geodesic of a photon arriving at $r=r_O$ at time 
$t=t_O$: the blue solid line is the numerical solution, the red dashed 
line 
is the analytical approximation of~(\ref{tL(r)}), and the black 
dotted line is the FLRW result.
In this plot $r_O=36.5 L$.
}
}

\end{figure}

We have also solved numerically ~(\ref{t-radial}), using the solution for 
$R(r,t)$ that comes from a numerical solution of the Einstein 
equations.
We compare the analytical approximation with the numerical result in 
figure~\ref{tempo}.

We note that the agreement in the plot is very good, but in general there can be a mismatch when $E(r)$ is non-negligible (for example, very far from the center). The same mismatch is found for $z(r)$ and $D_L(r)$. Remarkably, however, the mismatch disappears when one compares two physical quantities ($D_L-z$), confirming our expectation that a linear term in $E(r)$ should not have any physical effect.

%%%%%%%%%%%%%%%%%%%%%%%%%%%%%%%%%%%%%
\subsection{Redshift}
In the previous subsection we have found $t(r)$ along a photon 
trajectory. 
In this subsection we obtain the redshift $z(r)$ corresponding to a 
source located at $r$. The differential equation governing this 
relation 
is given by \cite{celerier}:
\be
{dz\over dr}={(1+z)\dot{R}'\over \sqrt{1+2E}} \, .
\ee
Again, ignoring $E(r)$, we were able to find an analytic 
approximation.
The result, derived in Appendix~\ref{app-redshift} can be written as:

\be
1+z(r)={t_O^{2/3}\over t_F^{2/3}}\exp\LT 
-\frac{2\epsilon}{ \pi}  \left( \frac{L}{r_{\rm 
hor}} \right) 
\cos\LF{2\pi r\over L}\RF
\RT \, , \label{zapprox}
\ee
Notice that the oscillating correction factor with respect to FLRW is once again proportional to the amplitude of the density fluctuations, $\e$ and to the scale $L$ divided by the horizon size, as in ({\ref{tL(r)}}). Here however the correction factor is  4 times larger, and also inside an exponential.

\begin{figure}
\psfrag{r}[][]{$(r-r_O)/h$ Mpc}
  \psfrag{-}[][]{}
    \psfrag{rO}[][]{}
    \psfrag{delta}[][]{$\sqrt{\langle\delta^2\rangle}=0.5$}
\includegraphics[width=0.78\textwidth]{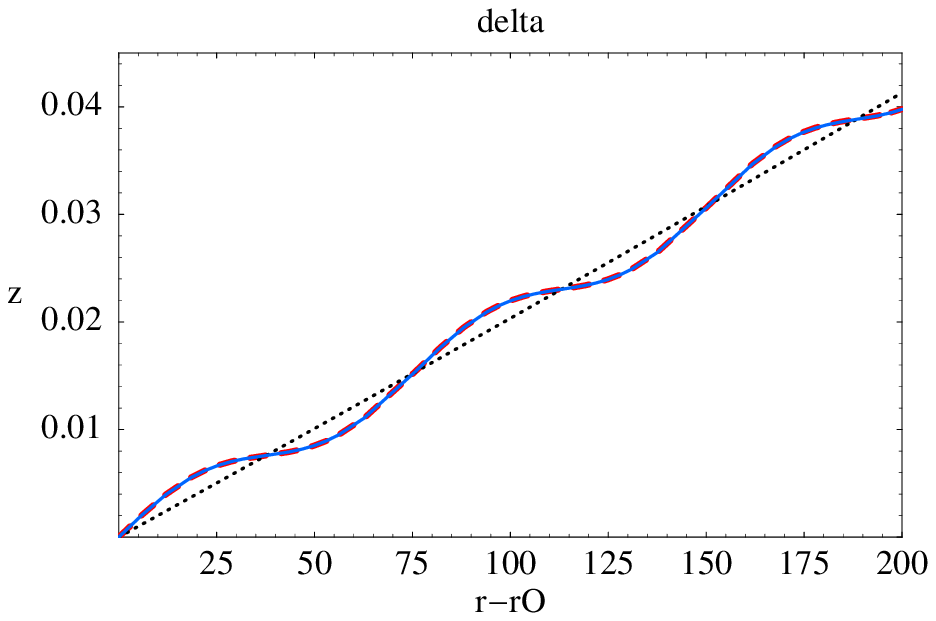}
\caption{\label{zetafig}
{\small
Redshift ($z$) along the geodesic of a photon arriving at $r=r_O$ at time 
$t=t_O$: the blue solid line is the numerical solution, the red dashed 
line 
is the analytical approximation, and the black 
dotted line is the FLRW result.
In this plot $r_O=36.5 L$, $t_O=3.3 \times 10^{17} {\rm 
sec}$.
}
}
\end{figure}

Again we have also solved the differential equation numerically and we 
compare the result with the analytic approximation in 
figure~\ref{zetafig}.

What  is very interesting to notice is that for small $z$ the correction is quite large. In fact expanding close to the observer we have
\be
z= {2\bb\bM \de r\over (\bM t_O)^{1/3}}\LT 1-{\e\over \pi} {L\over\de r}\cos\LF{2\pi 
 r \over L}\RF\RT \, ,
\label{z-linear}
\ee
where we have defined 
\be
\de r\equiv r-r_O \, .
\ee
One can check that the pre-factor corresponds to 
the FLRW relation. As we had alluded to before, the ``relative correction'' near the observer is controlled by the ratio $L/\de r$ instead of $L/r_{\mt{hor}}$, and this is one of the key results of the paper. Since near $t_O$, $\e\sim 1$, we see that in the first few oscillations when $\de r$
is close to $L$ the inhomogeneities  provide a large ${\cal O}(1)$ correction. This explains why several authors have obtained large corrections in the void models. 

%%%%%%%%%%%%%%%%%%%%%%%%%%%%%
\setcounter{equation}{0}
\section{Luminosity Distance} \label{luminosity}

In this section we show how to calculate the luminosity distance in our 
model.
As we said, we put the observer at a generic distance from the centre. Then we are able to calculate the luminosity distance as long as the source is aligned with the centre of the LTB model and the observer.

%%%%%%%%%%%%%%%%%%%%%%%%%%%%%%%%%%%%%%
\subsection{Off-centre observer}

The way we proceed is to notice first that in General Relativity there 
is an exact relationship between the so-called angular diameter 
distance ($D_A$) and the luminosity distance ($D_L$), and it is simply 
\cite{duality}:
\be
D_L=D_A (1+z)^2 \, .
\ee
So we focus on the calculation of the angular diameter distance.
It is well-known \cite{celerier} that for an observer at the centre of an LTB model the 
angular diameter distance is just:
\be
D_A=R|_{S} \, ,
\ee
where the subscript $S$ means that the quantity has to be evaluated at 
the space-time location of the source.
In order to find the distance $D_A$ for a generic location of the 
observer one has to solve the equation of motion for a bunch of 
geodesics 
infinitesimally close to the radial one. If the observer is not at the 
centre all these geodesics are not radial. Nonetheless we have been 
able 
to derive the exact form of $D_A$ in Appendix~\ref{DLapprox}, for radial sources. The final 
result is given by:
\be
D_A=R_S \left(R_O \int_{r_O}^{r_S} 
\frac{R'(r,t(r))}{(1+2 E(r))(1+z(r)) R(r,t(r))^2} dr \right)  \label{DLoffcentre} \, ,
\ee
where the integral has to be evaluated along the radial photon trajectory.

This expression can be easily evaluated numerically. 
One can get some insight into it by looking at the homogeneous limit (FLRW), by 
substituting $R=a(t) r$, and neglecting $E(r)$. 
In this case the correction factor inside the brackets is:
\be R_O\int_{r_O}^{r_S} \frac{R'(r,t(r))}{(1+2 E(r))(1+z(r)) 
R(r,t(r))^2} dr = 1-\frac{r_O}{r_S} \, .
\ee
This reduces to $1$ in the $r_O \to 0$ limit.
Moreover, for any $r_O$ this is what we expect for the FLRW limit, where
 the area distance is just
\be
D_A=a(t)(r_S-r_O) \, .
\ee

%%%%%%%%%%%%%%%%%%%%%%%%%%%%%%%%%%%%%%
\subsection{Analytical Estimate}
We observe that the $D_L-z$ relationship can {\it not} in 
general be expressed as a single-valued function $D_L(z)$.
The physical reason is that in some region of space (collapsing structure with 
high overdensity) the photon can get a blueshift. Therefore two 
points at different distances may have the same value of $z$ (similar to what was found in \cite{ellis2}).

\begin{figure}
\psfrag{r}[][]{$(r-r_O)/h$ Mpc}
  \psfrag{-}[][]{}
    \psfrag{rO}[][]{}
\psfrag{DL}[][]{$D_L$}
    \psfrag{delta}[][]{$\sqrt{\langle\delta^2\rangle}=0.5$}
\includegraphics[width=0.78\textwidth]{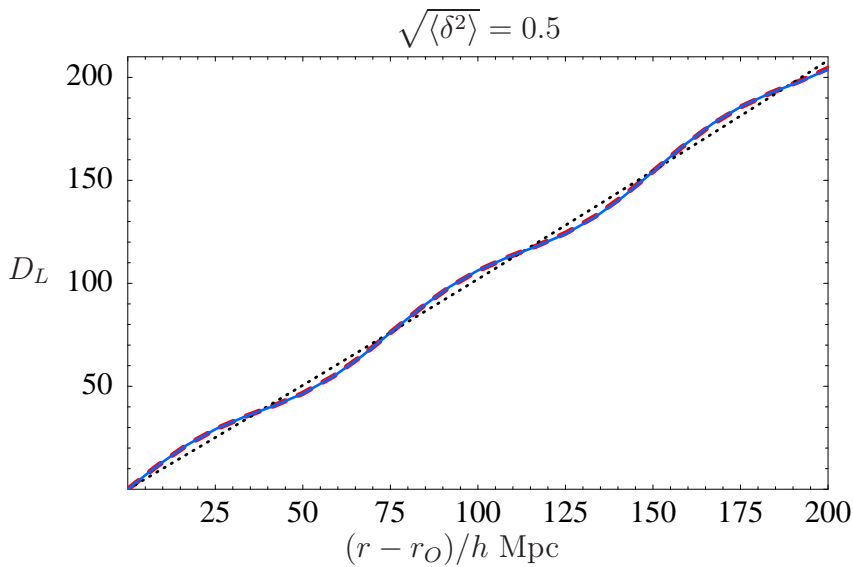}
\caption{\label{dielle-r}
{\small
Luminosity distance ($D_L$) along the geodesic of a photon arriving at $r=r_O$ at time 
$t=t_O$: the blue solid line is the numerical solution, the red dashed 
line 
is the analytical approximation, and the black 
dotted line is the FLRW result.
In this plot $r_O=36.5 L$, $t_O=3.3 \times 10^{17} {\rm 
sec}$.
}}
\end{figure}

So we have a parametric expression for $D_L-z$, where the parameter is 
$r$. In the previous section we have already obtained $z(r)$ (\ref{zapprox}).
The expression for $D_L(r)$ is derived in appendix~(\ref{DLapprox}):
\be
D_L(r)\approx 3\bb(1+z)\de r(\bM t_O)^{2/3}\LT 1-
\frac{\epsilon}{2 \pi} \left( \frac{L}{\de r} \right)
\cos\LF{2\pi  r\over L}\RF\RT
\label{D_L}  \, .
\ee
We compare this analytical approximation with a numerical solution in 
figure~\ref{dielle-r}.

\begin{figure}
\psfrag{r}[][]{$(r-r_O)/h$ Mpc}
  \psfrag{-}[][]{}
    \psfrag{rO}[][]{}

\psfrag{H0DL}[][]{$H_0 D_L$}
    \psfrag{delta}[][]{$\sqrt{\langle\delta^2\rangle}=0.5$}
\includegraphics[width=0.78\textwidth]{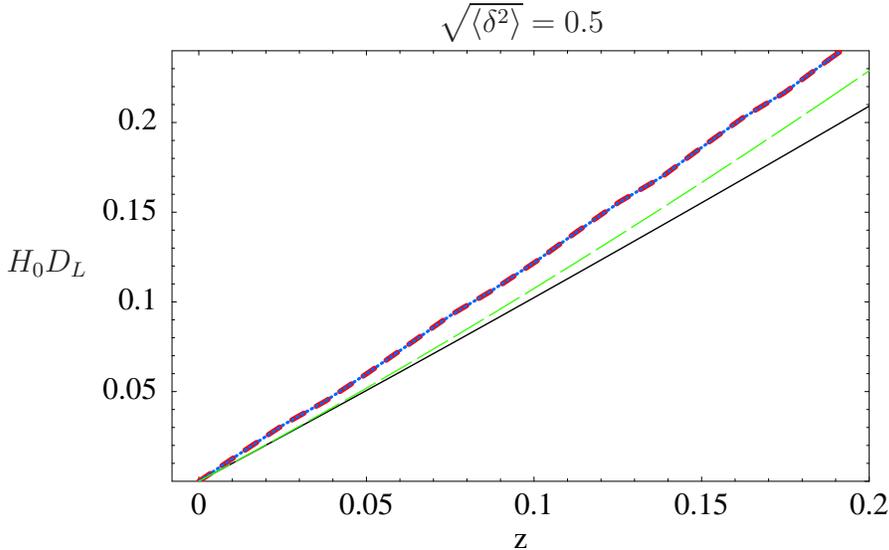}
\caption{\label{dielle-z}
{\small
Luminosity distance~($D_L$) vs. redshift~($z$): the blue solid line is the numerical solution, the red dashed 
line 
is the analytical approximation, the black 
dotted line is the EdS model ($\Omega_m=1$), and the  green long-dashed line is the 
$\Lambda$CDM 
result (with $\Omega_{\Lambda}=0.7$).
In this plot $r_O=36.5 L$, $t_O=3.3 \times 10^{17} {\rm 
sec}$ and the observer is in an underdense point.
}
}
\end{figure}

Again we comment on the size of the correction. Similar to what happens for $z$ in ({\ref{z-linear}}) the correction is large close to the observer (small $\delta r$ and $\epsilon$ of order 1).
Remarkably the correction in $D_L(r)$ is of the same kind as in $z(r)$ ({\ref{z-linear}}), but it is  half its size (close to the observer). So while considering the $D_L-z$ relationship the two effects will not cancel out, as one can see comparing eqs.~(\ref{z-linear}) and~(\ref{D_L}) and comparing also the numerical results of fig.~(\ref{zetafig}) and~(\ref{dielle-r}).
The net effect is shown in the parametric plot of fig.~(\ref{dielle-z}), where we compare the analytic and the numeric solutions for the $D_L-z$ relationship.

For large redshifts,  when the density contrasts are small, $D_L$ is a single valued function in $z$. For instance for large redshifts one can easily obtain $D_L(z)$:
\be
1+z\longrightarrow \LF {t_O\over t_F}\RF^{2/3}\mx{ and } D_L\longrightarrow 3\bb(1+z)^2(\bM t_F)^{2/3}\de r
%\label{z-der} \, ,
\ee
leading to
\be
D_L(z)=3t_O(1+z-\sqrt{1+z}) \, ,
\ee
which coincides with the flat matter dominated FLRW expression as expected, since at high $z$ the density contrast is very small.

%%%%%%%%%%%%%%%%%%%%%%%%%%%%%%%%%%%%%%%%%%%%%%%%%%%%%%%%%%%%%%%%%%
\setcounter{equation}{0}
\section{Application to Cosmology and ``Apparent'' Acceleration}   \label{apparent}
Now, we have all the computational tools that we need for applications to cosmology.
We stress first, that we have a cosmological model that allows to study nonlinear dynamics and light propagation, and this could be used to understand several phenomena (like gravitational lensing, or secondary CMB fluctuations).
However our focus in this paper is to try and understand better the hypothesis that an inhomogeneous Universe with only dust can mimick Dark Energy. 

To summarize, we have seen so far (see figs.~(\ref{tempo},\ref{zetafig},\ref{dielle-r})) that there is basically no overall effect on $D_L$ or on $z$ in our Onion model: all the quantities oscillate around the FLRW values, with the oscillation amplitude controlled by two parameters, $L/r_{\mt{hor}}$ and $\e$ (typically a product of these two parameters). The latter is expected to become of ${\cal O}(1)$ for scales in the range of tens of Mpc. However, for these scales the former ratio is too small,  $L/r_{\mt{hor}}\sim 0.01$, leading typically only to a correction of a few percent in the luminosity-redshift relationship. Note that a similar correction is present also in the evolution of the average matter density~\ref{density-avg}. Thus we find an overall effect of about a few percent coming from inhomogeneities. This is larger than what is computed using perturbative arguments~\cite{Huiseljak,rasanen,KMNR,frysiegel}, which give  a $10^{-5}$ effect, but still smaller than what is required to explain away dark energy. 

However, we have also seen that one cannot naively argue, based on the above, that inhomogeneous models will fail to give sizeable corrections (around 10\%) that is required to explain the supernova data or  any other significant effect\footnote{Such a conclusion was made by~\cite{Sugiura}, by studying an LTB model made of infinitely thin concentric shells.}. The oscillations in the $D_L-z$ relationship become a large effect close to the observer, because near the observer the relevant quantities, $z$ and $D_L$, are themselves small. We saw that the corrections do not proportionally become smaller but instead the relative corrections are now approximately governed by the ratio $L/\de r$. Thus in the first few oscillation cycles when $ L/\de r\sim {\cal O}(1)$ we in fact notice a sizeable correction (see fig.~(\ref{zetafig},\ref{dielle-r})). Based on our analysis we will now discuss three main issues:
\vs
1. How large is the effect due to a large scale fluctuation? Can it mimick Dark Energy? If it can, what is the relevant length scale $L$ and  amplitude $\e$? 
\vs
2. Can the fact that light travels through clumpy matter give rise to any interesting effects? Here we really try to go beyond the Onion model, but use it to estimate the corrections.
\vs
3. We discuss some observable deviations from the FLRW interpretation of the supernova data coming from inhomogeneities with ``typical'' amplitudes (\ie as predicted from CMB and/or observed in the galaxy power spectrum).

%%%%%%%%%%%%%%%%%%%%%%%%%%%%%%%%%%%%
\subsection{Can a large-scale fluctuation mimick Dark Energy?}  \label{largeA}
Here we explore the possibility that a fluctuation on a (large) scale $L$ could be responsible for a  very large effect on the $D_L-z$ relationship close to the observer at low and intermediate redshifts, such that it can mimick Dark Energy. The idea is that if we consider a very large scale, the local Hubble flow could be very different from the average one (indeed measurements at high redshifts give typically lower values of the Hubble constant, see the discussion in~\cite{HST}). We will consider the effect of one single wavelength for simplicity.
Then, we will vary the amplitude around (and above) the typical values inferred from the observed power spectrum. We will also vary the position of the observer with the same purpose of investigating if there is a setup that could mimick Dark Energy.
Crucially, we note  that the evidence for acceleration comes from the fact that we observe a mismatch between the expansion at low redshift (where the Hubble parameter is measured \cite{HST}, between roughly $0.03\leq z \leq 0.07$) and the expansion at higher redshift (where supernovae are observed~\cite{Riess}, between roughly  $0.4\leq z \leq 1$).  Indeed, we find that, depending upon the position of the observer (under-dense or over-dense) and the amplitude of density fluctuations, it is possible to account for this mismatch. 

In the following we will discuss how to account for (i) high redshift supernova data, (i) local measurements of the Hubble parameter,  (iii) local measurements of matter and baryon densities, and (iv) the  position of the first acoustic peak in CMB and as observed recently in matter power spectrum. We end with a  realistic assessment of this model and how one can discriminate this from the $\La$CDM one.

%%%%%%%%%%%%%%%%%%%%%%%%%%%%%%%%%%%%
\subsubsection{Fitting the high-z Supernovae}
For physical clarity it is useful to define an ``average'' model which is simply the homogeneous EdS model with the same value $t_O$ as the inhomogeneous one, or, equivalently, whose density goes as the average density in the Onion model:
\be
\langle \rho_{m}\rangle ={m_{Pl}^2\over 6\pi t^2}=\bar{\rho}_m \, ,
\ee
where we denote quantities corresponding to the homogeneous model with a ``bar''. The Hubble constant today ($\bar{H}_0$) in the average model is given by  
\be
\bar{H}_0={2\over 3t_O} \, .
\ee
Now, the important thing to realize is that the oscillations due to inhomogeneities for the scales that we are interested in become negligible before we reach the realm of high redshift supernovae, \ie  say for $ z\geq 0.4$. In this region therefore the luminosity distance roughly coincides with the EdS universe with a Hubble parameter $\bar{H}_0$:
\be
D_{E}={2\over \bar{H}_0}(1+z-\sqrt{1+z}) \, .
\label{D-EdS}
\ee

In order to match with observations one has to compare~(\ref{D-EdS}) with the observed redshift of supernovae. However, we know that, in this region, the  $\La$CDM model with $\Om_{\la}=0.7$ and $h=0.7$ is in good agreement with data. To understand analytically, it is  therefore sufficient to compare~(\ref{D-EdS}) with $D_L(z)$ in the ``concordance'' $\La$CDM model  given by
\be
D_{\La}={J_{\Om_{\la}}(z)\over 
H_0}\, , \qquad \mx{ where }J_{\Om_{\la}}(z)\equiv (1+z)\int_1^{1+z}{dy\over \sqrt{\Om_{\la}+y^3(1-\Om_{\la})  }} \, ,
\ee

So, we require that the ratio
\be
{D_{E}\over D_{\La}}=\LF{H_0\over \bar{H}_0}\RF{2(1+z-\sqrt{1+z})\over  J_{\Om_{\la}=0.7}(z)}\equiv  \LF{h\over \bar{h}}\RF{\cal D} \, ,
\ee
be roughly equal to one. It is important to note  that the ratio ${\cal D}$ changes very little in the relevant range,  $0.5\leq z\leq 1.5$: 
\be
{\cal D}(0.5)=0.83\geq {\cal D}\geq {\cal D}(1.5)=0.72 \, .
\label{curlyD}
\ee
Thus by choosing $\bar{H}_0$ appropriately: 
\be
0.83\geq {\bar{h}\over h }\geq 0.72  \, ,
\label{h-range}
\ee
we can be consistent with the high redshift supernova data. 

What is clear from~(\ref{h-range}) however, is that the Hubble constant of the average model has to be significantly smaller than what is typically measured, $h\sim 0.65-0.72$, by local observations, such as low redshift supernovae. In the rest of the paper, for analytical estimates, we will use $h=0.7$ which means that to be consistent with~(\ref{curlyD}) we need
\be
0.58\geq\bar{h}\geq 0.50 \, .
\ee
The success of the model therefore will rely on whether the local inhomogeneities can explain a larger local value of the Hubble parameter or not. This is what we discuss in the next subsection.
%%%%%%%%%%%%%%%%%%%%%%%%%%%%%%%%%%%%%%%%%%%%%%%%%
\subsubsection{Inhomogeneities \& Local Hubble constant}  \label{Hubble}
In order to see whether inhomogeneities can account for a larger local Hubble parameter let us first compute how $z$ is related to $D_L$, by combining eqs.
(\ref{z-linear}) and~(\ref{D_L}). As we pointed out earlier, the correction in the redshift close to the observer is  double the correction in $D_L$, therefore we have a net effect. For small $z$ we get
$$
D_L={3\over 2}\LF1+{\e\over 2\pi}{L\over\de r}\cos{2\pi 
 r \over L}\RF z \, t_O \, ,
$$
which is equivalent to
\be
D_L\approx {z\over \bar{H}_0}\LF1+{\e\over 2\pi}{L\over\de r}\cos{2\pi 
 r \over L}\RF = \frac{z}{H_0} \, ,
\ee
where we impose the last equality since we want to reproduce the local observations. Thus it is clear that, in order for our model to explain the measured Hubble parameter, we have to ensure, from ~(\ref{h-range})
\be
0.83\geq\LF1+{\e\over 2\pi}{L\over\de r}\cos{2\pi 
 r \over L}\RF={\bar{H}_0\over H_0}={\bar{h}\over h} \geq 0.72\, .
\ee
First of all we notice that one needs to consider a scale of fluctuation, $L$,  that at least extends up to $z\approx 0.07$ (corresponding to a $k\approx 1/65 \, h \, {\rm Mpc}^{-1}$), otherwise it is very difficult to ensure that the Hubble law  looks linear in the range
$0.03\leq z \leq 0.08$. This basically gives us an $L>200/h$ Mpc (in fact, we will see below that the optimal condition is $L/2\sim 200/h$ Mpc). Next let us try to estimate the density contrast that is necessary.

Our analytical formulas are only valid when the observer is located at a maximum or a minimum of the density profile, but this should be sufficient just  for the purpose of estimating the correction\footnote{We could vary the position of the observer using numerical solutions, as well, but it does not change the situation qualitatively.}. Let us consider these two cases separately. What we really want to find out is the maximal local Hubble parameter $h$ that we can obtain due to the inhomogeneities, given an $\bar{h}$.  Now, we get a maximal correction (in the right direction) when $\cos(2\pi  r / L)=-1$, in which case 
\be
{\bar{h}\over h}=\LF1-{\e\over 2\pi}{L\over\de r}\RF \, .
\ee
 
When the observer is at an overdense region, the smallest $r$ for which this happens is $\de r=3L/4$, and therefore to be consistent with~(\ref{h-range}) we find
\be
  \e\geq 0.8 \, .
\ee
On the other hand if the observer is at the minimal underdense region, we get a maximal correction at $\de r=L/4$ giving us a much larger correction and we only need
\be
 \e\geq 0.27 \, .
\ee
Thus, {\it prima facie},  it seems possible for the inhomogeneities to account for the  observed ``largeness''of the local Hubble parameter. The model prefers an underdense observer as one then requires smaller density contrasts and secondly for an overdense observer, the Hubble diagram is non-linear initially at very small redshifts $z<0.02$ which may still be conflict with observations\footnote{We have not carefully studied these very local observations because local measurements of matter  density almost certainly rules out having an overdense observer in the Onion model.}.

\begin{figure}
\hspace{-1cm}
\psfrag{delta}[][]{$\delta$}
\psfrag{z}[][]{$z$}
\psfrag{Lo}[][]{$5 Log_{10}D_L$}
\psfrag{H0DL}[][]{$H_{0} D_L$}
    \psfrag{Ldelta}[][]{$L=450/h$Mpc \, ; \, $\sqrt{\langle\delta^2\rangle}=0.34$}
\includegraphics[width=0.57\textwidth]{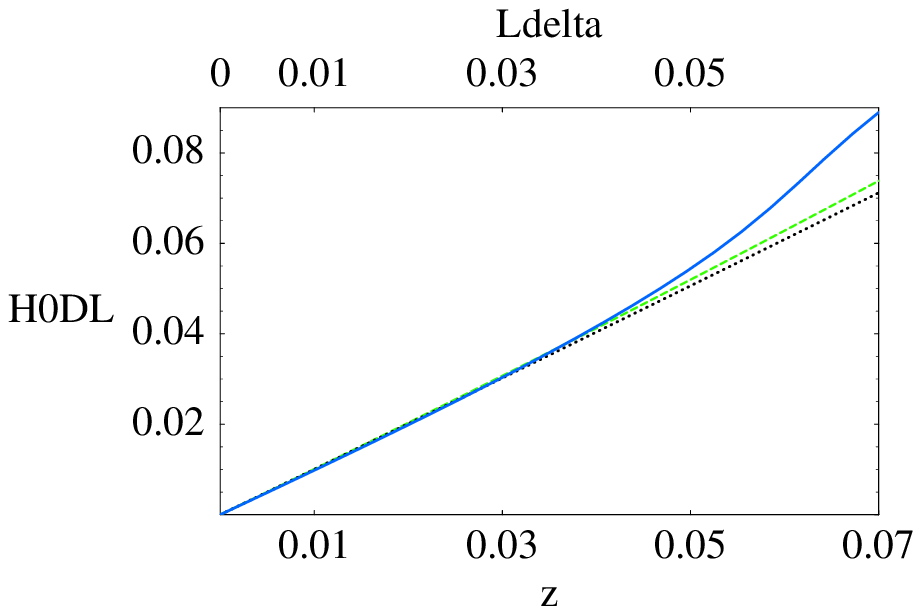}
\hspace{0.4cm}
\includegraphics[width=0.55\textwidth]{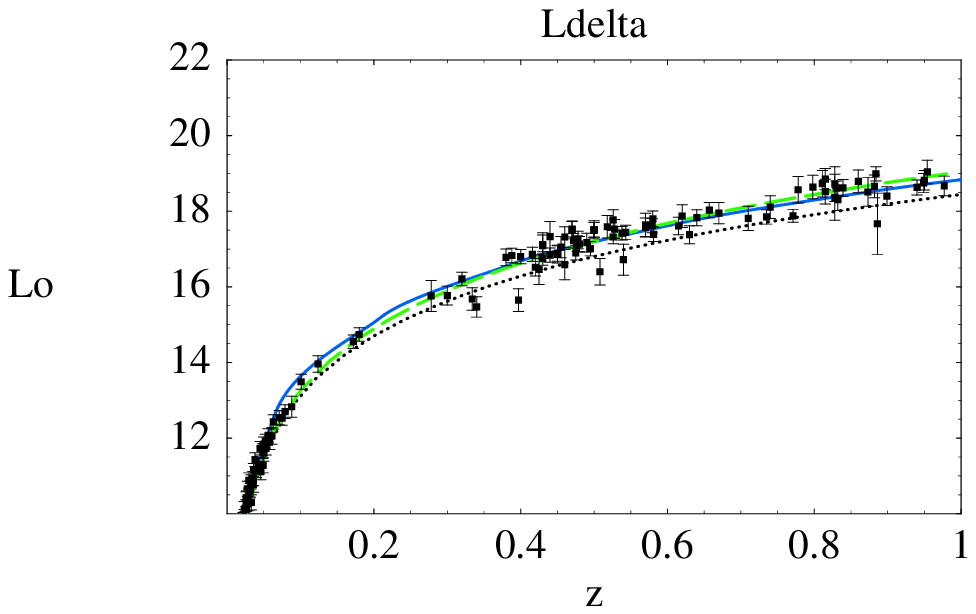}
\hspace{-2cm}
\includegraphics[width=0.5\textwidth]{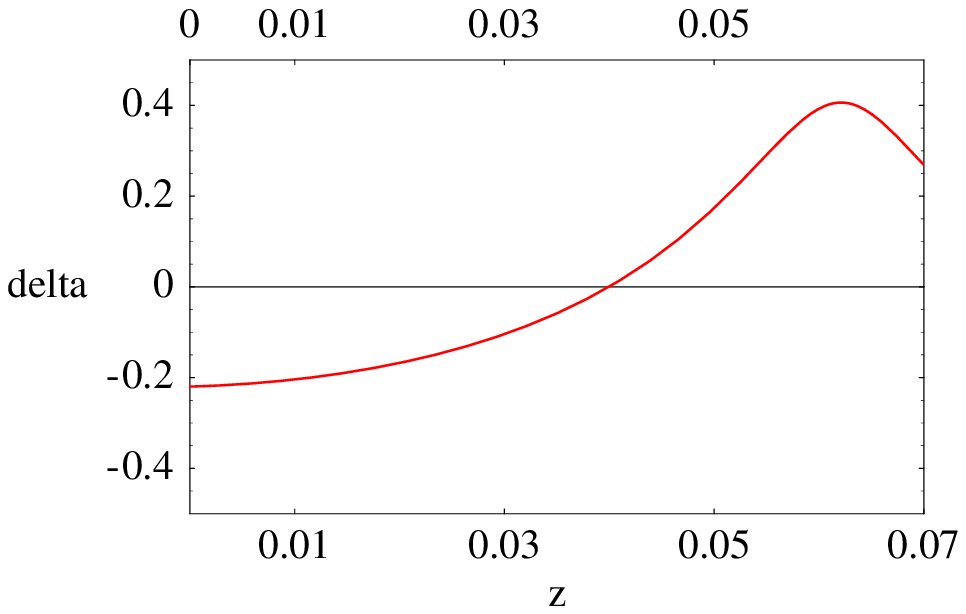}
\hspace{0.8cm}
\includegraphics[width=0.52\textwidth]{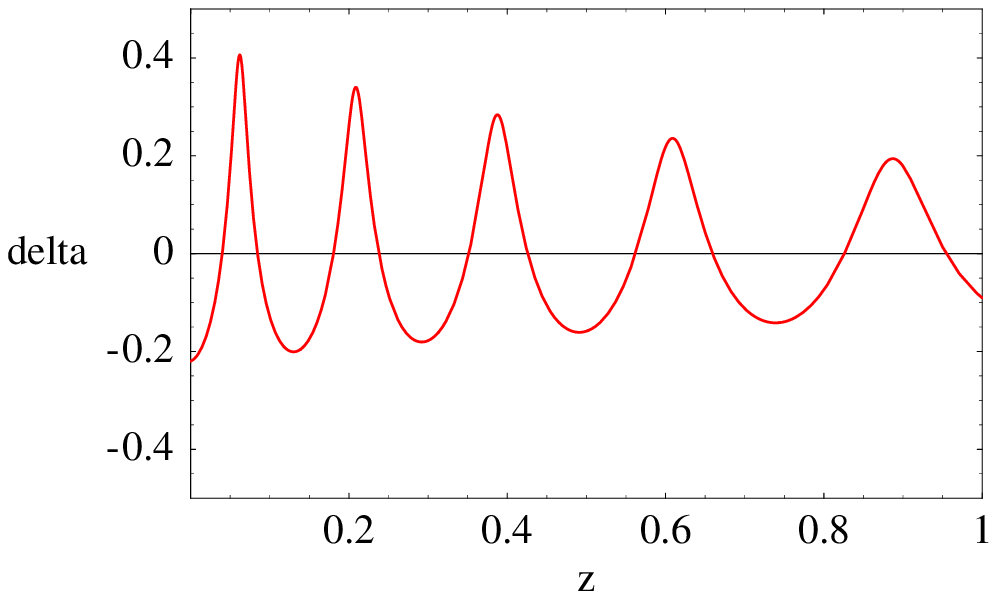}
\caption{\label{smallz}
{\small
Luminosity distance ($D_L$) vs. redshift ($z$) at very small (left) and intermediate (right) redshifts: the blue solid line is the numerical solution of
our model, 
the black dotted line is the EdS homogeneous model ($\Om_m=1$), the green dashed line is the $\Lambda$CDM model ($\Omega_{\Lambda}=0.73$). The models are normalized to have the same slope at low redshift.
In the right-hand side plot, we have superimposed the supernovae gold data set of~\cite{Riess}. 
The plots in the bottom show the density contrast (red line) as a function of $z$, for the model considered.
}
}
\end{figure}

\begin{figure}
\hspace{-1.2cm}
\psfrag{delta}[][]{$\delta$}
\psfrag{z}[][]{$z$}
\psfrag{deltam}[][]{$m-m_{\mt{empty}}$}
    \psfrag{Ldelta}[][]{$L=450/h$Mpc \, ; \, $\sqrt{\langle\delta^2\rangle}=0.34$}
\includegraphics*[width=1.1\textwidth]{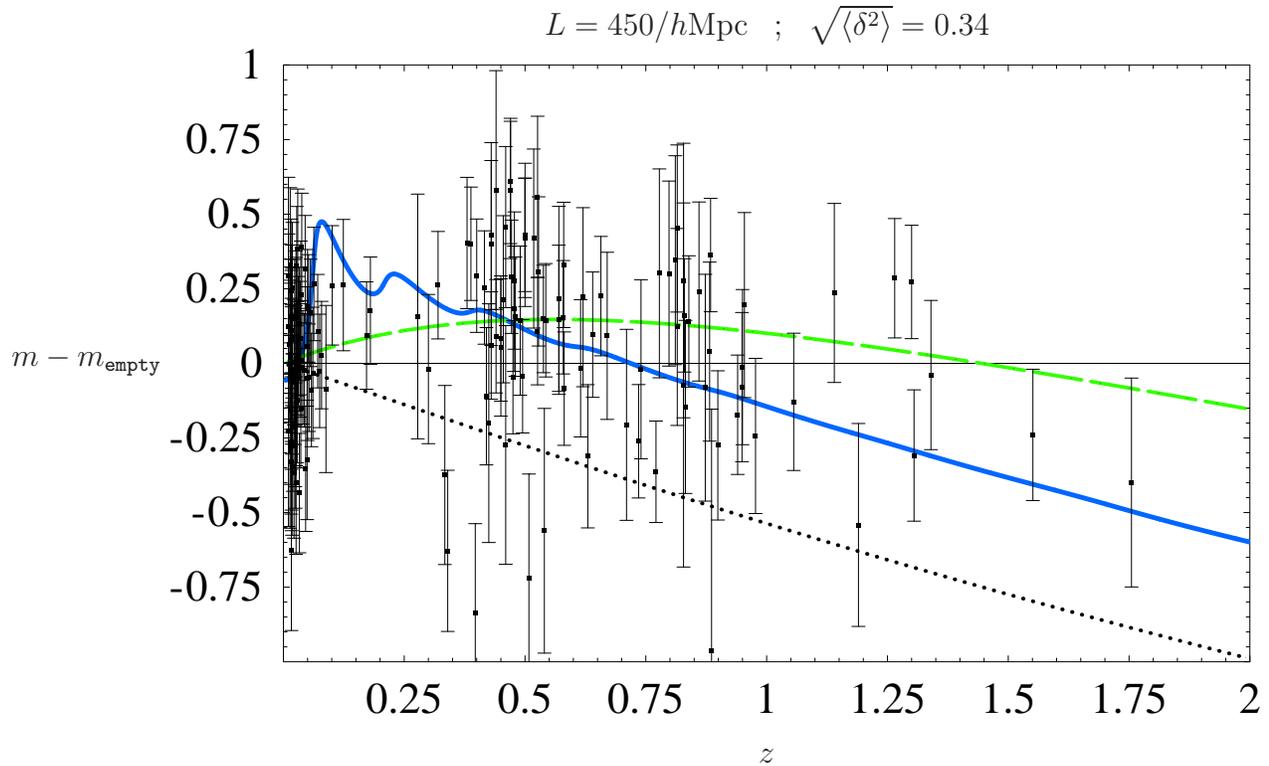}
\caption{\label{SN}
{\small
Magnitude residual from empty FLRW ($\Delta m$) vs. redshift ($z$): the blue solid line is 
our numerical solution, the black dotted line is the CDM 
model ($\Om_m=1$), and the green long-dashed line is the
$\Lambda$CDM 
result (with $\Omega_{\Lambda}=0.73$).
We have superimposed the supernovae gold data set of~\cite{Riess}.
}
}
\end{figure}

\begin{figure}
\psfrag{z}{$z$}
\psfrag{assey}[][]{$\frac{(m-m_{\mt{exp}})^2}{\sigma^2}$}
\includegraphics[width=0.8\textwidth]{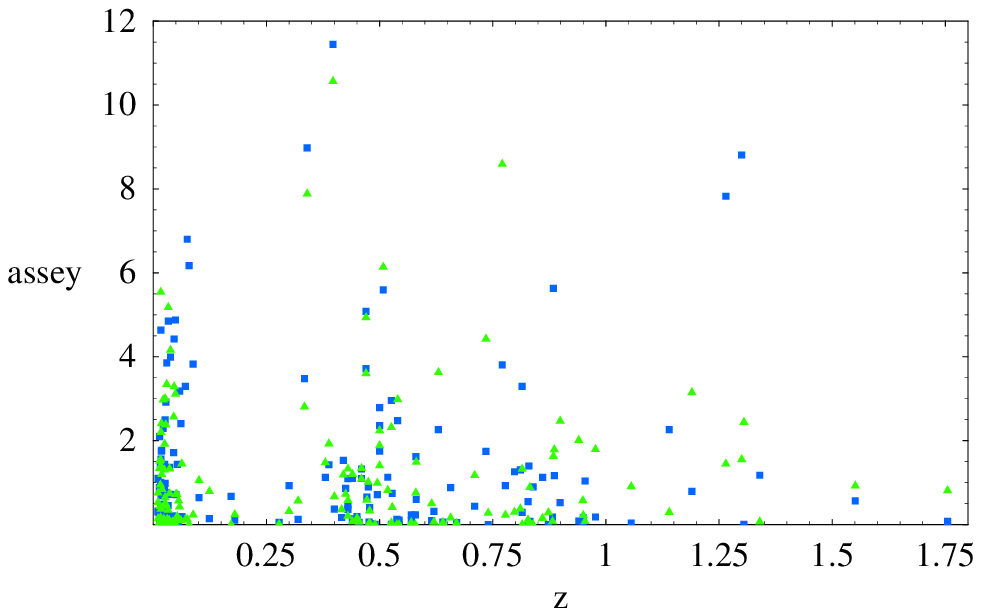}
\caption{\label{plotscarti}
{\small
We show here the contribution to the $\chi^2$ for each data point (gold data set of~\cite{Riess}) as a function of redshift, where $m$ is the theoretical apparent magnitude, $m_{\mt exp}$ is the observed magnitude and $\sigma$ is the experimental error. Our model (blue boxes) is compared with $\Lambda$CDM with $\Omega_{\Lambda}=0.73$ (green triangles).
}
}
\end{figure}

In fig.~(\ref{smallz}) we show one example, through numerical results, in which an almost linear $D_L-z$ relationship is recovered for $z\leq 0.07$, but a significant deviation from a CDM model is obtained at larger redshift, that reproduces something quite similar to a $\Lambda CDM$ model.
We show in fig.~(\ref{SN}) how the customary plot $m-m^{\rm empty}$ vs. $z$ looks like for these two models. In those plots  $m-m^{\rm empty}$ is defined as
\be
 m-m^{\rm empty}= 5 \log_{10}(D_L)-5 \log_{10}(D_L^{\rm empty}) \, ,
\ee
where $D_L^{\rm empty}$ is the luminosity distance for a homogeneous FLRW model with no matter and no cosmological constant, but just negative curvature. The reason for this definition is that a homogeneous model with no negative pressure components will always have $ m-m^{\rm empty}<0$. As one can see from the plots it is possible instead for an inhomogeneous model to have $ m-m^{\rm empty}>0$. 

Even though the model is not intended to be completely realistic (because of  having the extra spherical symmetry and only one wavelength), we can see from fig.~(\ref{SN}) that it can provide a reasonable fit to the experimental data.
We have performed also a $\chi^2$ analysis, and we have found the results given in tables~\ref{tabellaG},\ref{tabellaS}, that show explicitly that the model in the example gives a result not too far from the best-fit $\Lambda$CDM.
On the other hand from  fig.~(\ref{SN}) one can also see that the Onion model differs from the $\Lambda$CDM model in two regions: objects look fainter (larger $D_L$) at $0.1 \lesssim z \lesssim 0.35$ and look brighter (smaller $D_L$) at $z\gtrsim 1$.
With present data it is still hard to discriminate using this general feature: as one can see from fig.~(\ref{plotscarti}), individual points give similar contributions to the $\chi^2$, when comparing our Onion model and the $\Lambda$CDM model.

\begin{table}
\caption{Comparison with experimental data (gold data set of ~\cite{Riess})}
\vskip 0.2cm
\begin{tabular}{|l|r|}
\hline
Model&$\chi^2$ (157 d.o.f.)\\
\hline
$\Lambda$CDM homogeneous model (with $\Omega_M=0.27, \Omega_\Lambda=0.73$)& 178\\
EdS homogeneous model (with $\Omega_M=1.00, \Omega_\Lambda=0.00$) & 325 \\
Onion model ($\sqrt{\langle\delta^2\rangle}=0.34$ on $L=450/h {\rm Mpc}$ scale) & 212 \\
\hline
\end{tabular}
\label{tabellaG}
\end{table}

\vskip 0.7cm 
\begin{table}
\caption{Comparison with experimental data (full data set of ~\cite{Riess})}
\vskip 0.2cm
\begin{tabular}{|l|r|}
\hline
Model&$\chi^2$ (186 d.o.f.)\\
\hline
$\Lambda$CDM homogeneous model (with $\Omega_M=0.27, \Omega_\Lambda=0.73$)& 233\\
EdS homogeneous model (with $\Omega_M=1.00, \Omega_\Lambda=0.00$) & 403 \\
Onion model ($\sqrt{\langle\delta^2\rangle}=0.34$ on $L=450/h {\rm Mpc}$ scale) & 273 \\
\hline
\end{tabular}
\label{tabellaS}
\end{table}

To summarize, from our study we find surprisingly that we need a density contrast just slightly nonlinear ($\delta\simeq-0.3$) to account for the observed Hubble diagram. Nonetheless such a value represents a quite significant fluctuation about the observed average $\delta$ that is observed on this relevant scale $(|\delta|\simeq 0.05 - 0.1)$. Thus it is  fair to say that it is difficult (but perhaps possible) to explain away dark energy solely by including corrections coming from the first oscillation in the Onion model. Although the results are encouraging, probably it indicates that one has to go beyond the Onion model and take into account of other effects such as coming from the fact that light  travels mostly through voids, and this is what  we will discuss in the subsection~\ref{conj}. 

%%%%%%%%%%%%%%%%%%%%%%%%%%%%%%%%
\subsubsection{Matter density \&  CMB acoustic peak}
\label{1stpeak}
Having checked that  the Onion model can be consistent with  supernovae data and the Hubble diagram, we turn our attention to  other cosmological observations.
\vs 
{\bf Matter Abundance:} There are different cosmological observations (for example the measurement of the matter-radiation equality scale looking at the matter power spectrum) which measure the  matter density in our  local universe and our model has to be consistent with these measurements.  We already know that the $\La$CDM model with $h=0.7$ and $\Om_m\approx 0.3$, \ie with
\be
\rho_{\mt{measured}}={3\over 8\pi}\Om_m H_0^2m_{Pl}^2\simeq 4 \times 
10^{-47} {\rm 
GeV}^4 \, ,
\label{rhoobs}
\ee
provides a good fit to all these observations. On the other hand in our model the average density is given by 
\be
\bar{\rho}_{0}={3\over 8\pi} \bar{H}^2_0m_{Pl}^2 \, ,
\ee
so that
\be
0.58\leq {\rho_{\mt{measured}}\over \bar{\rho}_{0}}={\Om_m H_0^2\over \bar{H}^2_0}={\Om_m h^2\over \bar{h}^2}\leq 0.44
\label{rho-range} \, ,
\ee
where in obtaining  the  inequalities we have used~(\ref{h-range}). If we identify $\rho_{\mt{measured}}$ with the local value of matter density, this means that it has to be approximately half of the average value. In particular this implies that the we (observer) must be located in an underdense region. 

In fact one can make an estimate of the measured local density in our model. Typically a measurement of this sort involves measuring the mass in a certain region and then measuring the volume of the same region. The crucial point to note is that while measuring the volume, one uses the luminosity distance as the yard-stick of measuring distances. Thus, for the purpose of illustration,  if we consider a spherical region of radius $\de r$ around the observer, then the measured density would be  given by
\be
\rho_{\mt{measured}}={M\over V}={{4\over 3}\pi M_0^4 \de r^3\over {4\over 3}\pi D_L^3}=M_0^4\LF{ \de r\over D_L}\RF^3
\label{local} \, .
\ee 
Below we provide some numerical results for this quantity, but let us first try to estimate it analytically.

Substituting~(\ref{D_L}) in~(\ref{local}) we  find
$$
\rho_{\mt{measured}}={ M_0^4\over 6\pi (1+z)^6 (\bM t_F)^2 \LT1-{\e\over 2\pi}\LF{L\over \de r}\RF\cos {2\pi r\over L}\RT^3} \, .
$$
Or, approximately
\be
\rho_{\mt{measured}}={ \bar{\rho}_0\over\LT1-{\e\over 2\pi}\LF{L\over \de r}\RF\cos {2\pi r\over L}\RT^3} \, .
\ee
As an estimate let us consider the region upto the maximal correction (see previous subsection), so that $\cos(2\pi r/L)=-1$ at $\de r =L/4$, and we have
\be
\rho_{\mt{measured}}={ \bar{\rho}_0\over\LT1+{2\e\over \pi}\RT^3} \, .
\ee
In order to be consistent with~(\ref{rho-range}) we therefore find that we  need 
\be
\e\geq 0.48 \, .
\ee

The example in fig.~(\ref{smallz}-\ref{SN})  have a local density measured inside a radius of about $100$ Mpc of $1.8 \rho_{m,0}$. We leave a more detailed investigation of checking the consistency of observed matter densities coming from different measurements with the expectations  from the Onion models for future~\cite{BMN2}.
\vs
{\bf First acoustic peak in CMB:} We have seen how our model can be consistent with locally observed matter density by virtue of ``us'' living in an underdense region. However, the measurements of CMB are sensitive to only the average values since  our model reduces to the EdS model, already by a redshift of $z\sim 0.3$. This fact also ensures that our model can fit very well, at least the first peak position 
\be
l_1=l_1(\bar{\Om}_m,\bar{h},\bar{\Om}_b/\bar{\Om}_m,\bar{\Om}_{\ga}/\bar{\Om}_m)
\label{first-peak} \, ,
\ee
in CMB. In~(\ref{first-peak}) $b$ and $\ga$ stands for baryons and photons respectively. For instance, using $\bar{\Om}_m=0.9$, $\bar{h}=0.58$, and a standard value for the baryon-to-matter ratio, $\bar{\Om}_b/\bar{\Om}_m=0.13$, we find that CMBFAST \cite{CMBfast} gives good agreement to the first peak position. 
It would be necessary to see if we can be consistent with the full CMB spectrum. Although this is beyond the scope of the present paper, we notice that it is similar in spirit to what the authors of \cite{Sarkar} have done: by using a lower value of $h$ they could fit the CMB with an EdS model.
\vs
{\bf Baryon Oscillations:} Finally our model is consistent with the recent measurement of the baryon acoustic peak in the galaxy distribution \cite{Eisenstein}. In fact the angular diameter distance at $z=0.35$ looks very similar to the distance in the $\Lambda CDM$ model (with $\Omega_{\Lambda}=0.7$). Numerically we get for example the following values:
\begin{eqnarray}
D_A(0.35)&=&1375 \, {\rm Mpc} \qquad {\rm for \, \, \Lambda CDM } \, ,\\
D_A(0.35)&=&1386 \, {\rm Mpc} \qquad {\rm for \, Onion \, model \, with}\, \sqrt{\langle\delta\rangle^2}=0.34 \, . 
\end{eqnarray}
It is actually very hard to compare with the measured values reported in \cite{Eisenstein}:
\be
D_V(0.35)=1370 \pm 64 \, {\rm Mpc} \, ,
\ee
since the ``distance'' $D_V$ reported in the paper does not have a definition valid in general, but it is an FLRW based definition: $D_V\equiv (D_A^2 \, z/H(z))^{1/3}$, where $H(z)$ is the FLRW Hubble parameter.
However, since our results are very close to $\Lambda CDM$, it is clear that they  will be consistent also with the observations.

%%%%%%%%%%%%%%%%%%%%%%%%%%%%%%%%%%%%%%%%%%%%%%%%%%%
\subsubsection{Assessment} 
We comment here on the viability of the model, that we have explored:

1. As we have shown, the amplitude of the density contrast needs to be quite large ($\sim 0.3$) on a scale of roughly $400/h$ Mpc, with respect to the average values taken from the power spectrum ($0.05-0.1$). 

2. In this case, one also needs to ensure that the observer sits in a particular position that guarantees to have corrections of the same size, looking in different directions. This forces the observer to sit close to the minimum of a three dimensional valley (that is, close to the centre of a void).
This possibility  has already been proposed by several authors~\cite{celerier,pm84,kl92,Tomita00,Tomita01,wiltshire05,moffat,alnes,abtt06,m05,ps06} , and
we basically confirm the results of these previous papers, with the difference that we have a  model that is valid from the early to the late Universe and that it is not limited to the study of one void.

 3. Many observations would need to be fit carefully using our parameters: for example one should check if a value of the matter density a little higher than the observed value can be consistent with all the observations. And also if the baryon and the matter density can be made consistent with the full CMB spectrum (as we said, this has been shown to be possible under some additional assumption on the power spectrum by \cite{Sarkar}).

4. It will be possible to discriminate between this Onion model and the $\Lambda$CDM concordance model with more Supernova data: in the former model objects look fainter (larger $D_L$) at $0.1 \lesssim z \lesssim 0.35$ and look brighter (smaller $D_L$) at $z\gtrsim 1$ than in the latter.

Anyway the Onion model cannot be used directly as a fully realistic model, to interpret the data, since some crucial effects are hidden by the symmetries of the model. We discuss some of these effects in the next subsection.
%%%%%%%%%%%%%%%%%%%%%%%%%%%%%%%%%%%%%%%%%%
\subsection{Light propagation through voids} \label{conj}
Here we try to go beyond and compensate for a couple of crucial features missing in our Onion model. First, in the real world the overdensities virialize; instead in  LTB models  nothing prevents the thin shells to collapse to infinite density. This means we cannot follow the system beyond the time of collapse (so we can consider nonlinear effects only at very low redshifts). Moreover, due to virialization,  voids and overdensities behave very differently  from each other in the nonlinear stage.
Overdense regions are prevented from collapse by angular momentum.
There is nothing instead, that prevents a void to become more and more empty.
In our model there is an extremely delicate balance in the $D_L-z$ relationship between the effect given by voids and the effect given by structures.
A photon that travels through a  void gets significantly more redshift than the average, but this is canceled by the less redshift (or even blueshift!) that it gets when travelling through a collapsing region with high density.
Now, in the real universe it is unlikely that the collapse is so strong to give significant blueshift. So, taking into account of this fact might well break the cancellation between voids and structures. Note that, if this cancellation is broken, this will go exactly in the direction of mimicking an accelerating Universe \cite{BMN2}. In fact, a photon travelling in the late time Universe would get more redshift than a photon travelling in the ``linear'' Universe. As a result distant objects would look less redshifted than close ones (as in a homogeneous $\Lambda$CDM model).

The second limitation, which probably is the most relevant and interesting, 
is the fact that in the Onion model a light ray that comes from a 
distant source {\it has} to travel unavoidably through an 
equal number of underdense and overdense structures. This is {\it not} the case instead 
in the real Universe, where most likely a photon will never encounter 
nonlinear high density regions. In fact, only as long as the density fluctuations are linear, the fraction of volume in the Universe in underdensities is equal to the fraction in overdensities, but as they start to become nonlinear, the volume in voids overcomes the volume in structures. In this case, a photon most likely travels through voids and therefore typically there is no compensation between redshift in structures and voids. Here we only present an estimate of the possible effect, leaving a more accurate study to \cite{BMN2}.  

In this context we note that in the previous literature although people have tried to account for changes in luminosities due to this effect, and in fact found in some cases significant corrections, relatively little work has been done  to understand the effects on redshift.  One of the earliest attempts dates back to Kantowski~\cite{kantowski} who tried to estimate the corrections in a Swiss-Cheese model of the universe (an exact solution, in which spherical Schwarzschild ``holes'' with a mass in the centre are embedded in an FLRW universe), making use of the optical scalar equations~\cite{sachs} that regulate the evolution of $D_A$ . He  estimated in this model the correction on the redshift to be negligible, but found important corrections coming from $D_A$ to the deceleration parameter (up to $50\%$, see also~\cite{kantowski2}).
Afterwards Dyer and Roeder~\cite{dyerroeder}  applied the same optical scalar equation to a beam of light travelling in empty space (re-deriving the result by Zeldovich) and generalizing it to partially empty beams, confirming important corrections to the deceleration parameter. However, Weinberg~\cite{weinberg} has shown that even if a beam of light gets a significant correction to the distance while traveling in empty space, another beam  traveling close to a compact object gets an equal and opposite correction. Therefore on average the FLRW result is claimed to be correct, and this is interpreted as a consequence of the conservation of the number of photons.
While this statement seems correct (the total luminosity being proportional to the number of photons), it does not take into account variations on the redshift of the single photons (Weinberg's statement has been challenged in~\cite{ellis2}: if the underlying gravitational dynamics differs from FLRW, the use of the homogeneous $D_L-z$ is not justified in spite of the fact that the photon number is conserved). Weinberg's statement is often assumed to imply that with sufficiently many sources there in no need to worry about inhomogeneities.
However, even accepting the statement, the question remains whether with a finite number of sources we may observe a result very different from the average. 
More recently, some authors~\cite{lensing} have studied for this purpose the magnification and demagnification effect using lensing techniques (therefore ignoring again the effect on the redshift of light due to inhomogeneous expansion rate) and numerical N-body simulations claiming that the dispersion  in a realistic Universe is at most of a few percent, well below the intrinsic dispersion of Supernovae (and the effect can become important only at very high $z$). However, contrary to these observations in our model we have seen (both numerically and analytically)  that the correction in $z(r)$ is  significant  and must be properly taken into account while deriving the luminosity vs. redshift relation. 

\begin{figure}
\psfrag{r}[][]{$(r-r_O)/h$ Mpc}
  \psfrag{-}[][]{}
    \psfrag{rO}[][]{}
\psfrag{DL}[][]{$D_L$}
    \psfrag{delta}[][]{$\sqrt{\langle\delta^2\rangle}=10$}
\includegraphics[width=0.78\textwidth]{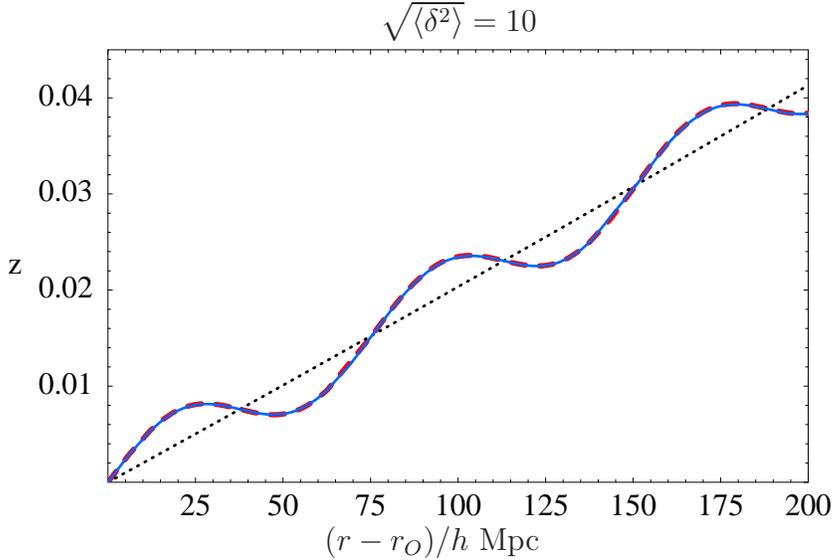}
\caption{\label{zetafig2}
{\small
Redshift($z$) along the geodesic of a photon arriving at $r=r_O$ at 
time $t=t_O$ in a strongly nonlinear regime: the blue dotted line is the numerical solution, the red 
dashed line is the analytical approximation~(\ref{zapprox}), and 
the 
black solid line is the FLRW result. 
In this plot $r_O=36 L$, $t_O=3.3 \times 10^{17} {\rm 
sec}$.
}
}
\end{figure}

In fact the correction in $z(r)$ is approximately double that of in $D_L(r)$  and importantly in the opposite direction. So in our attempt to try to estimate the effect of ``photons travelling mostly through voids''  in $D_L(z)$, we are only going to try to estimate the correction to the redshift in a given cycle, had it not passed through an overdense region, and then divide it by half. As one can see from a $z(r)$ plot (figure~\ref{zetafig2}), as the photon approaches a highly overdense structure, it first suffers from a blue shift (let's call the coordinate range through which this happens $\De r$) and then takes some distance (approximately equal to $\De r$) to recover from this blue shift. Thus as an estimate we see that the photon receives a ``zero-shift'' in a coordinate range $2\De r$. Thus it is natural to conjecture that if the photon had not encountered the structure it would not have lost the  $2\De r$ coordinate distance and therefore the increment in the redshift over a cycle should be corrected as
\be
(\De z)_{\mt{corr}}\approx (\De z)_{\mt{Onion}}\LF 1+2{\De r\over L}\RF=(\De z)_{\mt{FLRW}}\LF 1+2{\De r\over L}\RF \, .
\ee
The second equality follows from the fact that in the Onion model over a period, \ie length $L$, the change in the redshift is the same as the homogeneous EdS  model.

We note that this effect on the redshift (applied to CMB photons) has been estimated already by Zeldovich and Sazhin~\cite{Zeldovich2} in the context of a Swiss-cheese model.

Let us see whether we can estimate the crucial ratio $ \De r/ L$. It is clear that the blue-shift begins at a maximum of $z(r)$ and it ends at the minimum of the curve. Starting from the expression of $z(r)$ in~(\ref{zapprox}) one finds
\be
{dz\over dr}=0\Ra \e\LT 2\sin\LF {2\pi r\over L}\RF+{L\over \pi r_{\mt{hor}}}\cos\LF {2\pi r\over L}\RF\RT=1 \, .
\ee
The second term in the right hand side is much suppressed because of the ratio $L/r_{\mt{hor}}$ and therefore we approximately have
\be
\sin\LF {2\pi r\over L}\RF={1\over 2\e} \, .
\ee
This has periodic solutions signalling maxima and minima with an interval of 
\be
{2\De r\over L}=1-{2\over \pi}\sin^{-1}\LF{1\over 2\e}\RF \, .
\ee
As we mentioned before, the correction is really expected to be half of the above due to the cancelling effect in $D_L(r)$ and therefore we find
\be
\mt{Correction}\sim {\De r\over L}={1\over 2}-{1\over \pi}\sin^{-1}\LF{1\over 2\e}\RF \, .
\ee

For $\e= 1$, we get a maximal correction of  $1/3\sim {\cal O}(1)$, while  as we go back in time, the density contrast decreases and so does the correction. In fact, we find no blue-shift at all once $\e<0.5$, which occurs approximately at the redshift
\be
1+z={1\over \e_{\mt{min}}}\Ra z\sim 1 \, ,
\ee
precisely what is required to explain the high-redshift supernovae! 

In general if we consider a small scale, that becomes nonlinear earlier, this already can give some effect at higher $z$.

%%%%%%%%%%%%%%%%%%%%%%%%%%%%%%%
\subsection{Minimal uncertainties in the $D_L-z$ plot} \label{realistic}
In this section we discuss what are the {\it minimal} implications of our results on the distance-redshift measurements. From what we have seen so far,  it is perhaps possible that the effects are large enough to mimick Dark Energy, but it can also be that dark energy really is the explanation for the $D_L-z$ relation.
Even in this case, it is of great importance to study what are the deviations from FLRW.

We find two main effects: first, a generic observer will see a correction (at all redshifts) to $D_L-z$ of roughly $0.15$ apparent magnitudes with respect to the naive FLRW result, second any observer will see more scatter in the low redshift supernovae than in the high redshift ones.

To show this, we just take realistic values of the density contrast $\delta$ on some specific scale.
For simplicity we just use the same setup previously described, but we decrease the amplitude to the measured average value. To be more precise one would need to include fluctuations on all scales, but we leave this for a subsequent paper \cite{BMN2}.
So, we consider here an amplitude $\sqrt{\langle\delta^2\rangle}\approx 0.075$ (where $\langle ... \rangle$ is an average over many domains) on  a scale $L\approx 400/h {\rm Mpc}$, that is roughly consistent with the measured matter power spectrum.
We choose one particular position of the observer (underdense) and we compute numerically the deviations from the FLRW case.
We find uncertainties of order $0.15$ apparent magnitudes\footnote{A correction of this order was noted by~\cite{bolejko}, but it was not recognized that the corrections can affect all measurements, not just the low-redshift ones.}, which is the same size as the uncertainty on the intrinsic luminosity of a single Supernova (\cite{Riess}). 
Crucially, however, while the uncertainty in the intrinsic luminosity can be decreased by increasing the number $N$ of measured supernovae (the error decreases as $1/\sqrt{N}$), our result does not decrease since it is a real physical effect that depends on the mass distribution around us and it is therefore the most relevant systematic effect (for a discussion of the uncertainties see for example~\cite{hui}).

\begin{figure}
\hspace{-1cm}
\psfrag{delta}[][]{$\delta$}
\psfrag{z}[][]{$z$}
\psfrag{lo}[][]{$5 Log_{10}D_L$}
\psfrag{H0DL}[][]{$H_{0} D_L$}
    \psfrag{Ldelta}[][]{$L=400/h$Mpc \, ; \, $\sqrt{\langle\delta^2\rangle}=0.075$}
\includegraphics[width=0.55\textwidth]{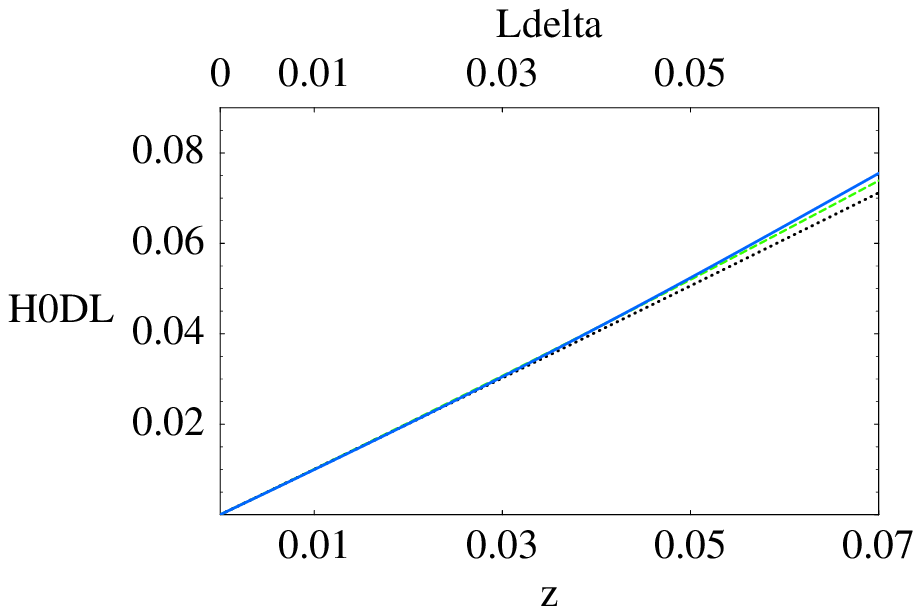}
\includegraphics[width=0.59\textwidth]{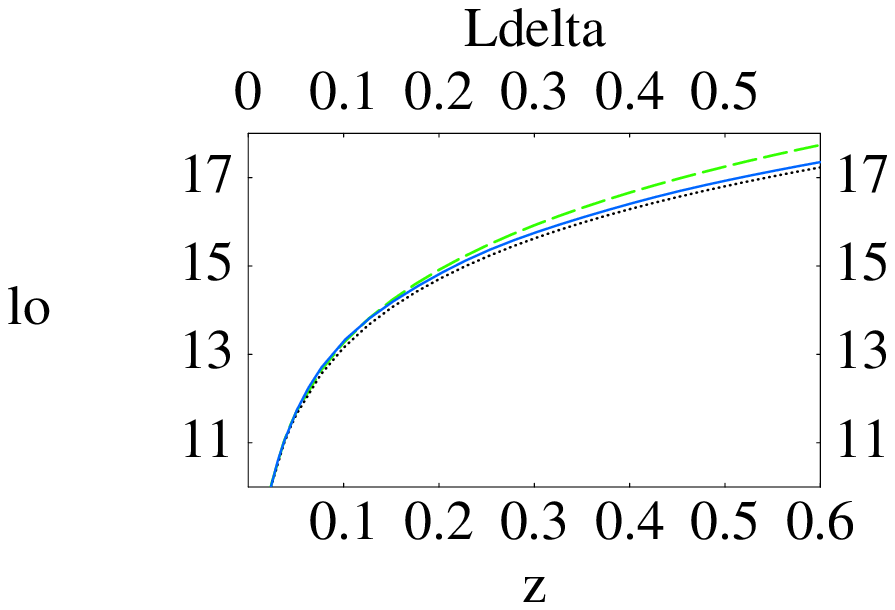}
\hspace{-1.8cm}
\includegraphics[width=0.5\textwidth]{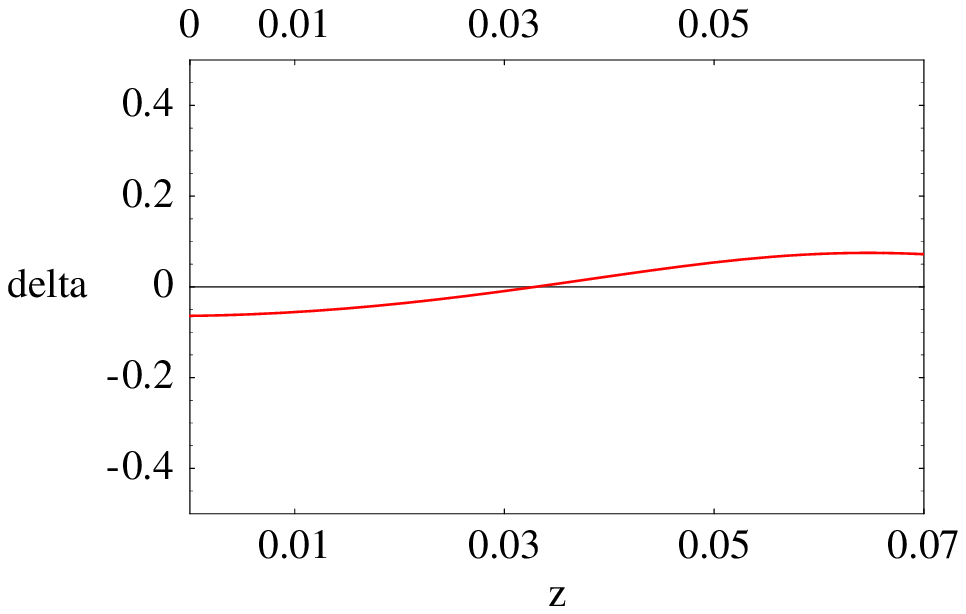}
\hspace{1.1cm}
\includegraphics[width=0.45\textwidth]{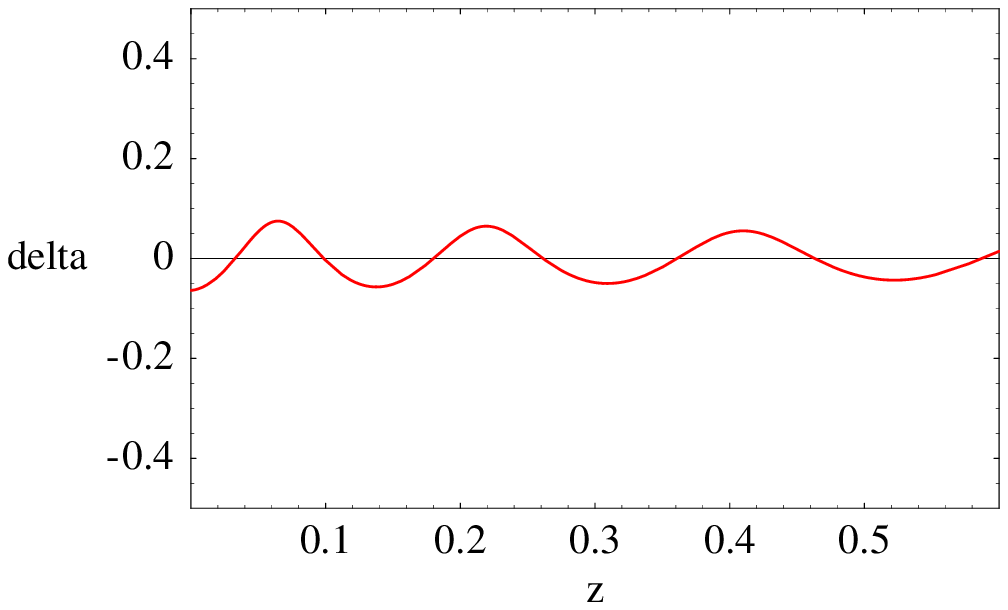}
\caption{\label{smallzr}
{\small
Luminosity distance ($D_L$) vs. redshift ($z$) at very small (left) and intermediate (right) redshifts: the blue solid line is the numerical solution of
our model, 
the black dotted line is the EdS homogeneous model ($\Om_m=1$), the green dashed line is the $\Lambda$CDM model ($\Omega_{\Lambda}=0.73$). The models are normalized to have the same slope at low redshift. 
The plots in the bottom show the density contrast (red dashed line) as a function of $z$, for the model considered.
}
}
\end{figure}

\begin{figure}
\hspace{0.8cm}
 \psfrag{deltam}[][]{$m-m_{\mt{empty}}$}
    \psfrag{Ldelta}[][]{$L=400/h$Mpc \, ; \, $\sqrt{\langle\delta^2\rangle}=0.075$}
\includegraphics[width=0.8\textwidth]{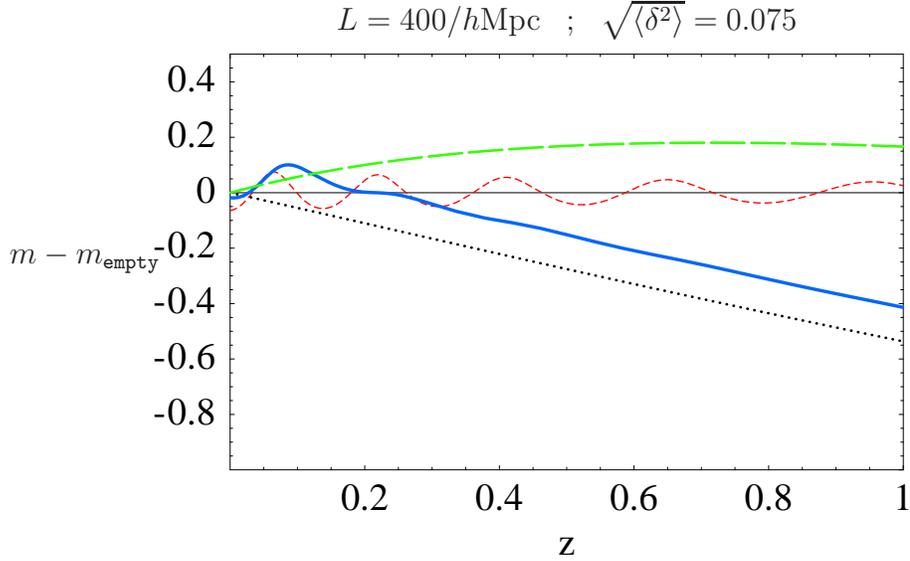}
\caption{\label{SNr}
{\small
Magnitude residual from empty FLRW ($\Delta m$) vs. redshift ($z$): the blue solid line is 
the numerical solution, the black dotted line is the CDM 
model ($\Om_m=1$), the green long-dashed line is the $\Lambda$CDM 
result (with $\Omega_{\Lambda}=0.73$).
We have superimposed the red thin dashed line, that shows the density contrast seen by the photon along its trajectory as a function of $z$.
}
}
\end{figure}

\begin{figure}
 \psfrag{deltam}[][]{$\Delta m$}
    \psfrag{Ldelta}[][]{$L=400/h$Mpc \, ; \, $\sqrt{\langle\delta^2\rangle}=0.075$}
\hspace{0.8cm}
\includegraphics[width=0.8\textwidth]{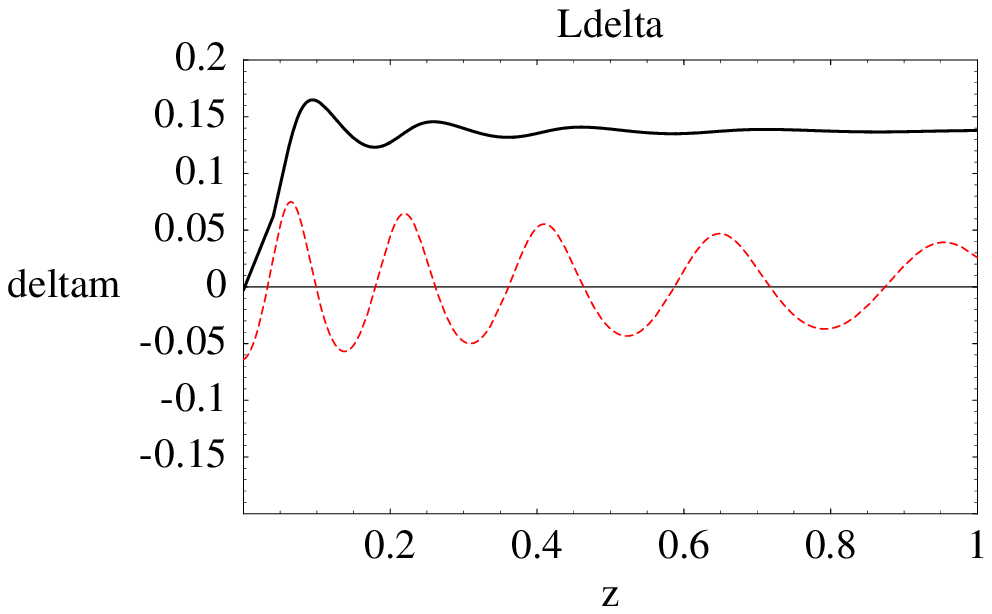}
\caption{\label{deltam}
{\small
The black solid line shows the correction in apparent magnitudes due to a realistic inhomogeneous density fluctuation ($\sqrt{\delta^2}\simeq 0.075$, for $L=400/h Mpc$) with respect to a homogeneous FLRW model.
We have superimposed the red thin dashed line, that shows the density contrast seen by the photon along its trajectory as a function of $z$.
}
}
\end{figure}

We show the ``Hubble'' diagram of the example analyzed in fig.~(\ref{smallzr}) and its magnitude compared to an empty universe in fig.~(\ref{SNr}). We show the ``error'' $\Delta m$ that is done using a CDM model with respect to this example in fig.~(\ref{deltam}) (we expect a similar result even in the presence of a dark energy component).

We note also that such an uncertainty is of the same order of the detected anisotropy of the Hubble constant at low $z$ \cite{anisotropy}.
It is usually assumed that the anisotropy is due to the velocity of our galaxy with respect to the FLRW frame (although with controversial results \cite{anisotropy}), but our results imply that the large scale structure plays a role that may be responsible for at least part of the anisotropy.
Similarly one could study the effect on the CMB due to the large scale structure (see for example \cite{Zeldovich2}, \cite{RasanenSchwarz}).

\begin{figure}
 \psfrag{deltam}[][]{$m-m_{\mt{empty}}$}
    \psfrag{Ldelta}[][]{$L=54/h$Mpc \, ; \, $\sqrt{\langle\delta^2\rangle}=1.6$}
\hspace{0.8cm}
\includegraphics[width=0.8\textwidth]{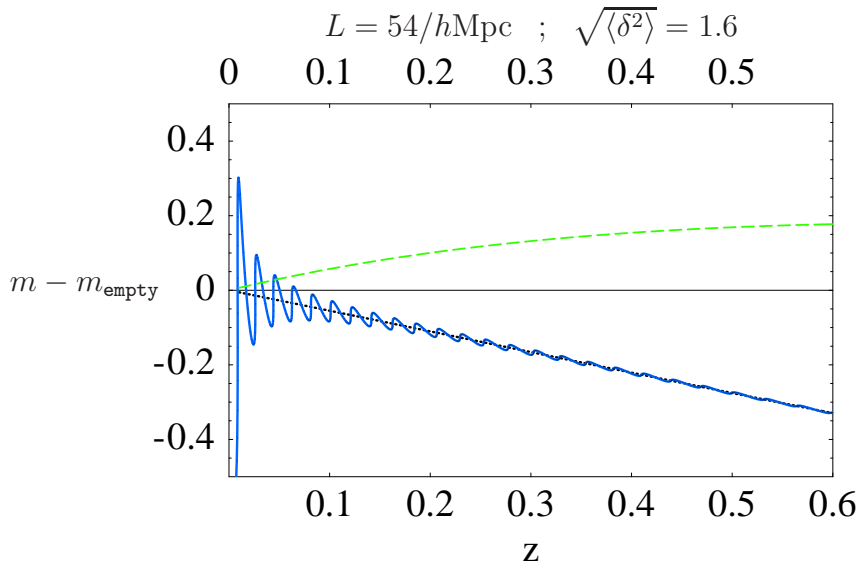}
\caption{\label{scatter}
{\small
The blue solid line shows the oscillations in the $D_L-z$ relationship with a $54/h$ Mpc nonlinear density fluctuation.
The black dotted line is the EdS model with the same high-z behaviour.
The green-dashed line, as usual, is the best-fit $\Lambda$CDM model ($\Omega_{\Lambda}=0.7$).
}
}
\end{figure}

Finally the other fact that we can account for is the experimental observation that low redshift supernovae show more scatter in apparent magnitudes than high redshift one~\cite{astier} (see also~\cite{bolejko}). As we have seen, in fact, the corrections in $z$ and in $D_L$ are larger close to the observer.
We show as an example a $D_L-z$ plot with a smaller wavelength (${\cal O}(50)/h Mpc$) in fig.~\ref{scatter}.

%%%%%%%%%%%%%%%%%%%%%%%%%%%%%%%%%%%%%%%%%%%%%%%%%%%%%%%%%%%%%%%%%%

\section{Conclusions}
\label{conclusions}

In this paper we have proposed an exact model, based on LTB metric with only dust, for structure formation applicable to a Universe homogeneous and isotropic on very large scale. 
The structures here look like concentric shells, and we could study their evolution from the early Universe to the present epoch.

In this background we have studied exactly the propagation of light along a radial trajectory, and we have computed the distance-redshift relation as measured by an observer sitting in a generic position.

For this purpose we have derived (for the first time) an expression for the luminosity distance in an LTB metric for an off-centre observer, we have solved the system numerically, and we have found accurate analytical approximations to the dynamical equations and to the light propagation equations.
We have made several conclusions on the basis of our computations.

First we notice that the corrections become larger close to the observer (far away the corrections due to underdensities tend to cancel the corrections due to overdensities). 

Then, we have investigated whether there is a setup that can mimick an accelerating $\Lambda CDM$ cosmology.
We have shown that this is possible under some special conditions: the observer has to be located around a minimum of the density contrast, and has to live in a large-scale underdensity with quite high density contrast: a typical example is $\delta\approx 0.3$ on a scale of roughly $400/h \, {\rm Mpc}$, while the typical value of $\delta$ at that scale is around $\delta\simeq 0.05 - 0.1$.

Without such a large density fluctuation the situation is still very interesting: one can estimate a typical effect, which affects all measurements, on the $D_L-z$ plot of about $\Delta m\simeq 0.15$ apparent magnitudes. By considering fluctuations of smaller length scales, $\sim 100/h$ Mpc, we can also explain the observation that low redshift supernovae should show more scatter than high redshift ones (as seen experimentally~\cite{astier}).

Finally, and most crucially, we notice that making the model more realistic could uncover relevant effects: the fact that light in a radial trajectory in the Onion model has to go through an equal number of structures and voids leads, in fact, to peculiar cancellations in the $D_L-z$ plot. Without these cancellations (which do not happen in the real world) we have estimated that the effect goes in the direction of apparent acceleration and that it should be large when matter fluctuations become nonlinear.
We leave a more detailed analysis of this effect to a subsequent paper~\cite{BMN2}.

\section*{Acknowledgements}

We are grateful to Gil Holder, Natalia Shuhmaher, Justin Khoury, Rashid Sunyaev, Syksy Rasanen, and Rocky Kolb for interesting discussions and suggestions. Tirtho would also like to thank the cosmology group at Tufts University for their hospitality during Tirtho's frequent visits to Boston.
The work of T.~B. and A.~N. is supported by NSERC.

%%%%%%%%%%%%%%%%%%%%%%%%%%%
\appendix
\setcounter{equation}{0}
\section{Validity of small $u$ and small $E$ approximation} 
\label{Esmall}
The analytical analysis in our paper relied on the $u$ variable being 
small. This is always true if one focuses on ``large'' enough values 
of $r$. Let us  now make this statement more precise.

In our model $E(r)$ is given by~(\ref{cosineE})
\be
E(r)={A_1\bM^2rL\over 2\pi}\left(1-\cos{2\pi 
r\over L}\right)={A_1\bM^2rL\over \pi}\sin^2\left({\pi r\over L}\right) \, .
\ee
Now,  since $E(r)$ is a periodic function, we have from~(\ref{v})
\be
v_{\mt{max}}\approx \left({9\sqrt{2}\over 
\pi}\right)^{\frac{1}{3}}(\bM t)^{1/3}\sqrt{{A_1L\over \pi r}} \, .
\ee
Thus the small $u$ approximation breaks down approximately at 
$r=r_u(t)$ given by
\be
r_u(t)={A_1L\over \pi}\left({9\sqrt{2}\over 
\pi}\right)^{\frac{2}{3}}(\bM t)^{\frac{2}{3}}\im{r_u\over L}=\left({9\sqrt{2}\over 
\pi}\right)^{\frac{2}{3}}{A_1\over \pi}(\bM 
t)^{\frac{2}{3}}
\label{r_u} \, .
\ee
From~(\ref{r_u}) it is clear that $r_u$ is a monotonically increasing 
function of time and since we are interested in describing the evolution 
only till today, we have
\be
{r_{u,\mt{max}}\over L}=\left({9\sqrt{2}\over 
\pi}\right)^{\frac{2}{3}}{A_1\over \pi}(\bM t_O)^{\frac{2}{3}} \, .
\ee
In other words, as long as $r>r_{u,\mt{max}}$, the small $u$ approximation is valid. We can 
actually estimate $r_{u,\mt{max}}$  more precisely by substituting $A_1$ from 
(\ref{A_1}):
\be
{r_{u,\mt{max}}\over L}=\de_C{\left({9\sqrt{2}\over 
\pi}\right)^{\frac{2}{3}}\over \pi \al}\left({t_O\over 
t_C}\right)^{\frac{2}{3}}= {20\over \pi }\left({t_O\over 
t_s}\right)^{\frac{2}{3}}\sim 6.4   \, .
\ee
Thus as long as our observer is located at a position $r_O> 6.4 L$, we can trust our small-$u$ approximation. In the numerical simulations we actually place the observer at a much larger distance, $r_O\sim 30 L$.

Let us now check when $E(r)$ remains small. Clearly this function grows with $r$ and therefore we have
\be
2E_{\mt{max}}=2A_1\left[{\bM^2(r_O+r_{\mt{hor}})L\over \pi}\right] \, ,
\ee
where $r_{\mt{hor}}$ is the coordinate interval that a CMB photon covers in its  journey towards us. Now, we can estimate this number using the flat FLRW 
model  (it is expected to change only slightly in the LTB model that we 
are considering). We find
\be
\bM r_{\mt{hor}}={(\bM t_O)^{1/3}\over \bb} \, ,
\ee
Thus we have
$$
2E_{\mt{max}}=2A_1(1+\vt){(\bM 
t_O)^{\frac{2}{3}}\over \bb^2}\left({L\over r_{\mt{hor}}\pi}\right)=(1+\vt)\left({9\over 
2\pi}\right)^{\frac{2}{3}}{\de_C\over \pi \al\bb^2}\left({L\over r_{\mt{hor}}}\right)\left({t_O\over 
t_c}\right)^{\frac{2}{3}} \, .
$$
Or,
\be
2E_{\mt{max}}={10(1+\vt)\over 3\pi}{L\over r_{\mt{hor}}}\left({t_O\over 
t_s}\right)^{\frac{2}{3}} \, ,
\ee
where we have defined
\be
\vt\equiv {r_O\over r_{\mt{hor}}} \, .
\ee

Since we are considering inhomogeneity scales roughly corresponding to 
400 Mpc/$h$, while the horizon is around 3Gpc/$h$, the ratio\footnote{ For large 
distances the ``proper distance'' $R\propto r$.} $L/ r_{\mt{hor}}\sim 0.1$ is not very small. In fact neglecting $E(r)$ introduces some mismatch between the analytical and numerical results in $D_L(r)$ and in $z(r)$. However, as we have discussed already in section~\ref{photon2}, the mismatch disappears when comparing physical observable quantities (such as $D_L$ vs. $z$). For smaller scales, instead, there is no mismatch at all.

%%%%%%%%%%%%%%%%%%%%%%%%%%%%%%%%%%%
\setcounter{equation}{0}
\section{Near-Centre Region} \label{nearcentre}
We have focused on $E(r)$ of the form~(\ref{cosineE}) because we want to describe the large scale periodic inhomogeneous structures. However, such an ansatz is actually inconsistent with the analytic requirements 
of $E(r)$ near $r=0$, and in particular in \cite{Flanagan} it was claimed  that this requirement does not allow for ``accelerating'' cosmologies. Since we put our observer at large distance away from the centre, it is perhaps obvious that this result is irrelevant for us, but here we provide one particular consistent $E(r)$ which for large $r$ reduces to that of the Onion model.  

It was pointed out in \cite{Flanagan} that one  requires the curvature to have 
the following form near $r=0$
\be
E(r)\sim r^n \mx{ with } n\geq 2 \, .
\ee 
 A simple function which has this property but asymptotes to~(\ref{cosineE}) for large $r$ is given by
\be
E(r)=\left({r^{n}\over r^{n-1}+b^{n-1}}\right)\left({A_1\bM^2L\over 
\pi}\right)\sin^2\left({\pi r\over L}\right) \, .
\ee
By construction 
\be
E(r)\longrightarrow\left\{\begin{array}{cc}{A_1\bM^2rL\over 
\pi}\sin^2\left({\pi r\over L}\right)& \mx{ for } r>b\\
  r^n\left({A_1\bM^2L\over \pi b^{n-1}}\right)\sin^2\left({\pi r\over 
L}\right)& \mx{ for } r<b
\end{array}\right. \, .
\ee

We note in passing that near the centre $u$ tends to increase ($u\ra \infty$ also corresponds to the curvature dominated regime). Thus we feel it is easy to attribute special properties to the central point and one of the reasons why we wanted to avoid placing the observer at the centre. 

It is possible to avoid $u$ becoming large by, for instance choosing  
\be
 {b\over L}>{20\over \pi}\left({t_O\over 
t_s}\right)^{\frac{2}{3}} \, ,
\ee 
but this also tends to wash-out the inhomogeneities.

%%%%%%%%%%%%%%%%%%%%%%%%%%%%%%%%%%%
\setcounter{equation}{0}
\section{The photon trajectory: $t(r)$} \label{tr}
We want to solve  the evolution equation for the photon trajectories 
given by~(\ref{t-radial}). As usual we ignore $E(r)$ in~(\ref{t-radial}). 
Then, substituting $R'$ we have for large $r$
\be
{dt(r)\over dr}=-3\bb (\bM t)^{\frac{2}{3}}\left[1+\al(\bM 
t)^{\frac{2}{3}}A(r)\right] \, .
\ee
It is convenient to work with the dimensionless variable
\be
x \equiv (\bM t)^{1/3} \, .
\ee
The differential equation then reads
\be
{dx(r)\over dr}=-\bb \bM\left[1+\al A_1x^2\sin\left({2\pi r\over L} \right) 
\right] \, ,
\label{dx-dr2}
\ee
Or,
\be
dx=-dr\ \bb\bM\left[1+\al A_1x^2\sin\left({2\pi r\over L}\right)\right] \, .
\label{dx-dr}
\ee

One cannot solve analytically this equation but we will adopt an 
 adiabatic approximation scheme where we treat the function $x(r)$ (or equivalently $t(r)$) to be slowly varying with respect to the oscillatory functions. Thus while differentiating or integrating, whenever we encounter a product of $x(r)$ (or one it's powers) with an oscillatory piece, we always neglect the variation of $x(r)$ and treat it as a constant. The physical reason why this becomes a good approximation is because while the variation in $x(r)$ is determined by the Hubble scale (inversely proportional), a variation in the oscillatory functions is inversely proportional to the coherence scale $L$. Thus as long as $L/r_{\mt{hor}}$ is small we are safe. 

Applying this method to the differential equation~(\ref{dx-dr}) we find
\be
x(r)=(\bM t_O)^{1/3}-\bb\bM(r-r_O)\left[1-{\al A_1L\over 2\pi(r-r_O)}x^2\cos\left( {2\pi 
r\over L}\right)\right] \, ,
\label{x(r)}
\ee
where we have assumed
\be
\cos\left( {2\pi r_O\over L}\right)=0 \, ,
\ee
\ie the observer is located either at a maximum or minimum of the 
density profile. First note, that we recover the FLRW limit by letting 
$A_1\ra 0$. In this case we find
\be
x(r)=x_O-\bb\bM(r-r_O)\equiv x_F(r) \label{xFLRW} \, ,
\ee
which is indeed the relation for an FLRW universe.
Let us now check explicitly that (\ref{x(r)}) is indeed a good approximate solution to the differential equation~(\ref{dx-dr}). 

Differentiating~(\ref{x(r)}) with respect to $r$ we obtain
$${dx(r)\over dr}\left[1-A_1{\bb \al\over  \pi}(\bM L)x\cos \left({2\pi 
r\over L}\right)\right]=-\bb\bM\left[1+\al A_1x^2\sin\left({2\pi r\over 
L}\right)\right] \, ,
$$
Comparing this with~(\ref{dx-dr}) we find that~(\ref{x(r)}) should be a 
good approximate solution provided the extra second term (relative to (\ref{dx-dr2}) is small:
\be
w\equiv A_1{\al\bb\over  \pi}(\bM L)x\ll 1 \, ,
\ee
Since, this is again a monotonic function in time we have
$$w_{\mt{max}}=\de_c{\bb\over  \pi}{(\bM L)(\bM t_O)^{1/3}\over (\bM 
t_c)^{\frac{2}{3}}} \, ,
$$
where we have further used~(\ref{A_1}). In order to estimate this 
quantity note that the horizon at a given $x$ can be estimated using just 
the FLRW result~(\ref{xFLRW}).
Using~(\ref{t_s}) we therefore find
\be w_{\mt{max}}={\de_c\over \pi}\lf{L\over r_{\mt{hor}}(x_O)}\rfi\lf {t_O\over 
t_c}\RF^{\frac{2}{3}}={1\over \pi}\lf{L\over r_{\mt{hor}}(x_O)}\RF\LF 
{t_O\over t_s}\RF^{\frac{2}{3}}\ll1  \, .
\ee
Thus,~(\ref{x(r)}) is indeed a good approximate solution.  

Solving~(\ref{x(r)}) we find
\be
x(r)={\left[1-\sqrt{1-4 \bb \al{x_F(r)A_1L\bM\over 2\pi}\cos\left( {2\pi 
r\over L}\right)}\right]\over 2 \bb \al{A_1L\bM\over 2\pi}\cos\left( {2\pi 
r\over L}\right)} \, ,
\label{xL(r)}
\ee
so that
\be
t(r)=\bM^{-1}x^3(r)
\label{t(r)} \, .
\ee

We have checked that~(\ref{t(r)}) is in excellent agreement with the 
numerical solution. In fact keeping upto linear order in $A_1$ in 
(\ref{xL(r)}) 
\be
x(r)\approx x_F(r)\LT 1+ \al A_1x_F(r){L\bM\over 
(6\pi)^{\frac{2}{3}}}\cos\LF {2\pi r\over L}\RF\RT
\label{x-approx} \, ,
\ee
we already get an excellent approximation. 

It is instructive to write the above 
expression  as 
$$
x(r)\approx x_F(r)\LT 1+ {\e\over  2\pi }\LF{L\over r_{\mt{hor}}}\RF\cos\LF {2\pi 
r\over L}\RF\RT \, ,
$$
which makes the dependence of the correction on the parameters of the model apparent.
%%%%%%%%%%%%%%%%%%%%%%%%%%%%%%%%%%%%%
\setcounter{equation}{0}
\section{Calculating the Redshift}\label{app-redshift}
In 
this appendix  we try and obtain the redshift $z(r)$ corresponding to a 
source located at $r$. The differential equation governing this 
relation is given by
\be
{dz\over dr}={(1+z)\dot{R}'\over \sqrt{1+2E}} \, .
\ee
Again, ignoring $E(r)$, we have 
$${dz\over 1+z}=\dot{R}' dr =2\bb\bM \LT 
x^{-1}(r)+2\al x(r)A(r)\RT dr
$$
$$\approx 2\bb\bM \LT x_F^{-1}(r)-{\bM L \al A_1\over 
(6\pi)^{\frac{2}{3}}}\sin{2\pi r\over L}+2\al x(r)A(r)\RT dr \, ,
$$
where we have used~(\ref{x-approx}). The first term can be integrated 
in a straightforward manner to yield the FLRW result. The second term 
can also be integrated easily. For the third integration one can use the 
same trick of first integrating with respect to $r$ assuming that $x$ 
is a constant and then put back the $r$-dependence of $x$.  After all 
these integrations we find
$$
\ln(1+z)-\ln C=-2\ln x_F(r)$$
\be 
-{(\bM L)^2\al A_1\over 3\pi (6\pi)^{1/3}}\sin\LF {2\pi r\over L}\RF 
-{2\over \pi}\al \bb A_1 \bM L x(r)\cos{2\pi r\over L} \, ,
\ee  
where $C$ is an integration constant. Now, the last two terms are only 
important close to $t_s\sim t_O$. During this time the second term is 
always a lot smaller than the third because $\bM L\ll x_s$ and thus we 
have
$$
\ln(1+z)-\ln C=-2\ln x_F(r) -{2\al\bb\over \pi} A_1 \bM L x(r)\cos{2\pi 
r\over L} \, .$$
We can fix the integration constant by demanding that at $r=r_O$, 
$z=0$. This gives us 
\be
\ln C=2\ln x_O \, ,
\ee
so that we have
\be
\ln(1+z)=-2\ln {x_F(r)\over x_O} -{2\al\bb\over \pi}A_1 \bM L x(r)\cos{2\pi 
r\over L} \, .
\ee
Or,
\begin{eqnarray}
1+z(r)\approx {x_O^2\over x_F^2(r)}\exp\LT -{2\al\bb\over \pi}A_1 
\bM L x_F(r)\cos\LF{2\pi r\over L}\RF\RT\approx \nonumber \\
\approx {x_O^2\over x_F^2(r)}\LT1- {2\al\bb\over \pi}A_1 
\bM L x_F(r)\cos\LF{2\pi r\over L}\RF\RT \, .
\label{z(r)}
\end{eqnarray}

Again, it is instructive to write the above expression as
$$1+z={x_O^2\over x_F^2(r)}\exp\LT -{2\e\over \pi}\LF{L\over 
r_{\mt{hor}}}\RF\cos\LF{2\pi r\over L}\RF\RT\approx{x_O^2\over x_F^2(r)}\LT1- {2\e\over \pi}\LF{L\over 
r_{\mt{hor}}}\RF\cos\LF{2\pi r\over L}\RF\RT \, .
$$
%%%%%%%%%%%%%%%%%%%%%%%%%%%%%%%%%%%%%%
\setcounter{equation}{0}
\section{$D_L$ for an off-centre observer} \label{OA}

In this appendix we derive an expression for the luminosity (or angular) distance as seen by an observer that looks in a radial direction and that sits in a generic off-centre position, at coordinate $r_O$.
In order to do that, we will need to consider trajectories that end up at the observer space-time position: however we need not only radial trajectories, but also those that have an infinitesimal deviation angle.
Then, given a solid angle $\delta\Omega$ at the observer for a bunch of light rays that arrive from the source, we will be able to compute the cross sectional area $A$ along the past light cone. This gives us the ``angular diameter distance'' (or the ``area distance'') $D_A=\frac{\delta \Omega}{A}$.
As we said, this is related immediately to the luminosity distance of a source by $D_A=(1+z)^2 D_L$.
Finally we will double-check our derivation by using the optical scalar equation derived by Sachs in 1961 \cite{sachs,kantowski} in appendix \ref{sachs}.

%%%%%%%%%%%%%%%%%%%%%%%%%%%%%%%%%%%%%
\subsection{Quasi-radial trajectories}

First of all, we give the geodesic equations that we will use to study the trajectories of the photons towards the observer.
The relevant geodesic 
equations are given by \cite{humphreys}:
\begin{eqnarray}
{d^2t\over dv^2}+{\dot{R}'R'\over 1+2E}\lf{dr\over 
dv}\rfi^2+\dot{R}R{\cal L}^2&=&0 \, ,
\label{t-evolution}  \\
{d^2\te\over dv^2}+2{R'\over R}{dr\over dv}{d\te\over 
dv}+2{\dot{R}\over R}{dt\over dv}{d\te\over dv}-\sin\te\cos\te\lf{d\phi\over 
dv}\rfi^2&=&0 
\label{theta-evolution}  \, , \\
{d^2\phi\over dv^2}+2{R'\over R}{dr\over dv}{d\phi\over 
dv}+2{\dot{R}\over R}{dt\over dv}{d\phi\over dv}+2\cot\te{d\te\over 
dv}{d\phi\over dv}&=&0  \, ,
\label{phi-evolution}
\end{eqnarray}

where $v$ is an affine parameter, and where we have dropped the equation for $r(v)$, since we will not need it.
In addition, the condition ($ds^2=0$) gives us:
\be
-\lf{dt\over dv}\rfi^2+{R^{'2}\over 1+2E}\lf{dr\over 
dv}\rfi^2+R^2{\cal L}^2=0   \, ,
\ee
where
\be
{\cal L}^2=\lf{d\te\over dv}\rfi^2+\sin^2\te\lf{d\phi\over dv}\rfi^2 \, .
\ee

Now, first we choose the observer to sit in the $\te=\pi/2$ plane. And then we observe that for symmetry reasons a trajectory lying in this plane will never escape it: we can in fact  consistently set $\te=\pi/2$ in the geodesic equations as all the terms 
in~(\ref{theta-evolution}) vanish. 
The axial symmetry of the system also ensures that it is sufficient to consider only such trajectories for the purpose of computing the luminosity distance.
So, (\ref{phi-evolution}) then simplifies to
\be
{d^2\phi\over dv^2}+2{R'\over R}{dr\over dv}{d\phi\over 
dv}+2{\dot{R}\over R}{dt\over dv}{d\phi\over dv}=0  \, .   \label{phiv}
\ee
Again,  without loss of generality we can assume the observer to lie along the ray  $\phi=0$. 
The radial trajectory will then be characterized by having $\phi(v)=0$.

However, in order to compute the luminosity distance, we need to find another ``near-by'' trajectory which, 
although reaches $(r=r_0,\phi=0)$ at $t=t_O$, starts, say at $t=t_i$, with 
$\phi= \phi_i$. The parameter $\phi_i$, in fact, parameterizes a bunch of 
geodesics which reach the same destination but diverge in the past. Now, 
we will only need to find trajectories which are infinitesimally close 
to the radial geodesic, \ie $\phi_i\ll 1$. This provides a great 
simplification, as we notice that ${\cal L}^2\sim \phi_i^2$ and therefore 
$r(\la,\phi_i)$ and  $t(v,\phi_i)$ do not get any correction $\sim {\cal 
O}(\phi_i)$. To see this, observe that in general for any coordinate
\be
x^{\mu}(v,\phi_i)= \left. x^{\mu}(v,0)+\phi_i {\p 
x^{\mu}(v,\phi_i)\over \p\phi_i}\right|_{\phi_i=0} \, .
\ee
For  $\phi(v,\phi_i)$  this means
$$\phi(v,\phi_i)= \left.\phi_i {\p\phi(v,\phi_i)\over 
\p\phi_i}\right|_{\phi_i=0} \, ,
$$
so that 
\be
{\cal L}^2=\lf{d\phi\over dv}\rfi^2=\left.\phi_i^2\lf {\p^2\phi\over 
\p v\p\phi_i}\right|_{\phi_i=0}\right)^2
\label{L} \, ,
\ee
Thus the equations~(\ref{t-evolution}) and~(\ref{L}) do not get any 
correction of the order $\sim \phi_i$ and therefore 
\be
r(v,\phi_i)=r(v,0)+{\cal O}(\phi_i^2)\mx{ and } 
t(v,\phi_i)=t(v,0)+{\cal O}(\phi_i^2) \, .
\ee

%%%%%%%%%%%%%%%%%%%%%%%%%%%%%%%%%%
\subsection{Off-centre observational coordinates}

Before calculating the luminosity distance we need to do an additional step: we define the system of coordinates used by the observer (the so-called ``observational coordinates''\cite{EllisRev}).
They are defined by the following conditions:
\begin{description}
\item[I.\ \ ] The time-like coordinate is the proper time of the observer 
\item[II.\ ]  The first spatial coordinate is a monotonic parameter along the past light cone: it can be an affine parameter ($v$) or the coordinate $r-r_O$ or anything else.
\item[III.]  The remaining two coordinates are ``angular coordinates'' $(\theta,\phi)$, in the sense that the metric along the light cone (keeping fixed the other two coordinates) looks like:
\be
ds^2= v^2 (d\tilde{\theta}^2+d\tilde{\phi}^2 \sin\tilde{\theta}^2)  \label{observational} \, .
\ee
\end{description}

For the time coordinate, we can directly use $t$, as it is already the proper time.
Then we attack the problem of finding the correct angular variables by first defining a rigid translation of the spatial coordinates, centering them in the observer, and then rescaling the angles in such a way that  ~(\ref{observational}) is satisfied.

Let us first concentrate on a coordinate transformation from the centre 
of the LTB patch, $C$, to the observer point located somewhere off the 
centre, $O$. The original LTB coordinates are called $x^{\mu} = \lbrace 
t, r, \theta, \phi \rbrace$.
We assume the observer to be comoving with the dust. In reality 
our Galaxy may have a peculiar velocity relative to the 
local LTB flow, but for the sake of simplicity we ignore it. 
That is, we assume the observer to be at a fixed coordinate 
distance $r_O$ from the centre. Now, let us define a new coordinate 
system with $O$ as the origin, denoted by $\tilde x^{\mu} = \lbrace \tilde t, \tilde r, \tilde 
\theta, \tilde \phi \rbrace$. The following relations hold between these two set 
of coordinates:
\begin{eqnarray}\label{coordtr}
\nonumber t &=& \tilde t\, , \\
\nonumber r \sin\theta \sin\phi &=& \tilde r \sin\tilde\theta 
\sin\tilde\phi\, , \\
\nonumber r \sin\theta \cos\phi &=& r_O + \tilde r \sin\tilde\theta 
\cos\tilde\phi\, , \\
r \cos\theta &=& \tilde r \cos\tilde\theta.
\end{eqnarray}
Without 
loss of generality we assume the three points $C, O,$ and $S$ to build our 
plane of $\theta = \tilde \theta = \frac{\pi}{2}$. The zero reference 
line of $\phi$ and $\tilde{\phi}$ in this plane is then defined by the line $C-O$. 

Our goal is now to  determine the metric in 
the new coordinates, using
\begin{equation}
g_{\tilde \mu \tilde\nu} = g_{\alpha\beta} \frac{\partial x^{\alpha}
}{\partial \tilde x^{\mu}} \frac{\partial x^{\beta}}{\partial \tilde 
x^{\nu}} \, .
\end{equation}   
The non-vanishing partial derivatives that we will need to calculate the area 
distance are as follows:
\begin{eqnarray}
\nonumber \frac{\partial r}{\partial \tilde \phi}\vert_{\theta = \tilde 
\theta = \frac{\pi}{2},\tilde \phi = 0} = 0 \, ,
\\
\nonumber \frac{\partial \theta}{\partial\tilde\theta}\vert_{\theta = 
\tilde \theta = \frac{\pi}{2},\tilde \phi = 0} = \frac{\tilde r}{r}\, , \\
\frac{\partial \phi}{\partial \tilde\phi}\vert_{\theta = \tilde \theta 
= \frac{\pi}{2},\tilde \phi = 0} = \frac{\tilde r}{r} \, .
\end{eqnarray}

The angular part of the metric at the observations point looks like:
\be
ds^2|_{t,\tilde{r}=const}= R^2 \frac{\tilde{r}^2}{r^2} (d\tilde{\theta}^2+d\tilde{\phi}^2 \sin\tilde{\theta}^2) \, .
\ee
At this point, in order to satisfy  ~(\ref{observational}) we redefine coordinates by:
\begin{eqnarray}\label{barred}
\nonumber \bar{t} &=& \tilde t\, , \\
\nonumber \bar{r} &=& \tilde r \, , \\
\nonumber K \bar{\phi}  &=& \tilde\phi \\
K \bar{\theta} &=& \tilde\theta +(K-1) \frac{\pi}{2} \, .
\end{eqnarray}
In this way the angular part of the metric looks like (around $\theta=\pi/2$):
\be
ds^2|_{t,\bar{r}=const}= K^2 R^2  \frac{\bar{r}^2}{r^2} (d\bar{\theta}^2+d\bar{\phi}^2) \, .
\ee
So we require
\be
K=lim_{v \to 0} \left(\frac{r \, v}{\bar{r} R }\right)= \frac{r_O}{ R \frac{d\bar{r}}{dv}}  \, ,
\ee
where we have used the fact that\cite{humphreys, EllisRev}:
\be
\frac{dt}{dv}=-(1+z) \, .  \label{vt}
\ee
In this way the barred coordinates are observational coordinates.

Finally, we will make use of the transformation between angles in the two coordinates systems: 
\be
\phi=\arccos \left[\frac{r_O + \bar{r}\,\cos (K \bar{\phi} 
)}{{\sqrt{r_O^2 + \bar{r}^2 + 2\,r_O\,\bar{r}\,\cos (K \bar{\phi} )}}}   \right] 
\simeq \frac{\bar{r}}{\bar{r}+r_O} K \bar{\phi} \label{arccos} \, ,
\ee
where the last equality holds for infinitesimal angle.

%%%%%%%%%%%%%%%%%%%%%%%%%%%%%%%%%
\subsection{Calculating the luminosity distance}\label{LD}
By definition, the area distance is the ratio between the solid angle in the observational coordinate and the area along the past light-cone \cite{EllisRev}:
\be
D_A^2 \equiv \frac{dA}{d\Omega} = \frac{d\theta_S d\phi_S}
{d\bar{\theta}_O d\bar{\phi}_O}  \sqrt{g_{\theta \theta} g_{\phi \phi}} \, ,
\ee
where as usual $S$ stands for source and $O$ stands for observer, and where we have already chosen $\theta=\pi/2$. This can be rewritten as:
\be
D_A^2 = \frac{d\theta_S d\phi_S }{d\bar{\theta}_O d\bar{\phi}_O} \, 
R|_S \, .
\ee
We have therefore to compute the ratios
\be
\frac{d \theta_S}{d\bar{\theta}_O}   \, , \qquad  \frac{d\phi_S} 
{d\bar{\phi}_O}  \label{ratios} \, ,
\ee
for infinitesimal angle around $\bar{\phi}=0$ (which corresponds to 
aligned Centre, Source and Observer). By axial symmetry it is sufficient to consider one ratio: the two 
ratios are the same.
We will therefore consider the ratio
$
\frac{d\phi_S} {d\bar{\phi}_O}  \label{rapporto} \, .
$

In order to find what this ratio is, we consider the geodesic equation  ~(\ref{phiv}). 
Remarkably it can be simply written as:
\be
\frac{d}{dv} \left( R^2 \frac{d \phi}{dv} \right)=0 \, .
\ee
So, the following quantity is a constant along the geodesic:
\be
R^2 \frac{d \phi}{dv} = C    \label{vr} \, .
\ee
So we may integrate the equation as:
\be
\phi_S= C \int_O^S \frac{dv}{R^2} \, ,
\ee
where we have put $\phi=0$ at the observer.
We basically already have the result, since we have the final angle and so the final area $A$.
Explicitly, we may change variable of integration along the geodesic, 
instead of the affine parameter $v$.
Using  ~(\ref{vt}), we can express the integral as:
\be
\phi_S= C \int_{r_O}^{r_S} \frac{R'(r,t(r))}{(1+z(r)) \sqrt{1+2 
E(r)} R(r,t(r))^2} dr  \label{result} \, .
\ee
The last thing to be worked out is the constant $C$. Making use of  ~(\ref{arccos}) we have that:
\be
\frac{d\phi}{dv}=\frac{d\phi}{d\bar{r}} 
\frac{d\bar{r}}{dv}+\frac{d\phi}{d\bar{\phi}} \frac{d\bar{\phi}}{dv}=\frac{d\phi}{d\bar{r}} 
\frac{dr}{dt}\frac{dt}{dv}+\frac{d\phi}{d\bar{\phi}} 
\frac{d\bar{\phi}}{dv} \, .
\ee
In the last equality we have used the fact that at the first order it 
is sufficient to consider a radial trajectory, where $d\bar{r}=dr$.
Explicitly evaluating the derivatives at $\bar{r}=0$ gives:
\be
\left.\frac{d\phi}{d\bar{\phi}}\right\arrowvert_O=0  \, , \qquad  
\left.\frac{d\phi}{d\bar{r}} 
\frac{dr}{dt}\frac{dt}{dv}\right\arrowvert_O= \frac{\bar{\phi}_O}{R_O} \, ,
\ee
which means that $C$ has the value:
\be
C= R_O \bar{\phi}_O \, .
\ee
This allows us to calculate the ratio of~(\ref{ratios}):
\be
\frac{d\phi_S} {d\bar{\phi}_O}=R_O  \int_{r_O}^{r_S} \frac{R'(r,t(r))}{(1+z(r))\sqrt{1+2 E(r)}  
R(r,t(r))^2} dr \, .
\ee
So the final result luminosity distance $D_L=(1+z)^2 D$,  is given by:
\be
D_L=(1+z)^2 R_S \left( R_O  
\int_{r_O}^{r_S} \frac{R'(r,t(r))}{(1+z(r)) \sqrt{1+2 E(r)} R(r,t(r))^2} dr \right)
\label{dlresult} \, .
\ee
With a very good approximation we can ignore $E(r)$ with respect to 1:
\be
D_L=(1+z)^2 R_S R_O\left(  \int_{r_O}^{r_S} 
\frac{R'(r,t(r))}{(1+z(r))  R(r,t(r))^2} dr \right)\equiv (1+z)^2 R_S R_OI  \label{dlresultA} \, .
\ee
%%%%%%%%%%%%%%%%%%%%%%%%%%%%%%%%%%%%%%%%%%%%%%%%%%%%%%5
\subsection{Analytical Approximation}  \label{DLapprox}
In this section we will simplify the expression for luminosity distance~(\ref{dlresultA}) for the Onion model. In order to do this first we will evaluate approximately the integral $I$.

Using~(\ref{onionR}), (\ref{onionRp}) and~(\ref{z(r)}) we find
\be
{R'\over (1+z)R^2}\approx {1- {2\al\bb\over \pi}x_FA_1\bM L\cos\LF 2\pi r\over L\RF +\al A_1x_F^2\sin\LF 2\pi r\over L\RF\over 3\bb x_O^2r^2} \, .
\ee
It is easy to see that for the realistic $L$, the third term always dominates over the second and therefore can be ignored. Moreover while integrating the third term one can once again use the adiabatic approximation ($(x_F/r)^2$ is changing slowly as compared to the sine), so that we straight-forwardly get
\be
I={\de r\over 3\bb r r_O x_O^2}\LT 1-{\al A_1x_F^2Lr_O\over 2\pi r \de r}\cos \LF 2\pi r\over L\RF\RT \, .
\ee

Substituting $I$ in~(\ref{dlresult}) we get
$$D_L=(1+z)^2 R_S R_O{\de r\over 3\bb r r_O x_O^2}\LT 1- {\al A_1x_F^2Lr_O\over 2\pi r \de r}\cos \LF 2\pi r\over L\RF\RT
\, .
$$
Using the expression for $R$ (\ref{R}) in the small $u$ approximation 
we find
\begin{eqnarray}
D_L\approx 3\bb (1+z)^2\de rx_F^2\LT 1-{\al A_1x_F^2Lr_O\over 2\pi r \de r}\cos \LF 2\pi r\over L\RF\RT 
\approx \nonumber \\ 
\approx 3\bb (1+z)\de rx_O^2\LT 1- {\al A_1x_F^2Lr_O\over 2\pi r \de r}\cos \LF 2\pi r\over L\RF\RT 
\label{Dappr} \, ,
\end{eqnarray}
where in the last approximation we have used~(\ref{z(r)}). This gives us the luminosity distance as a function of $ r$. Coupled with~(\ref{z(r)}) we can obtain a parametric plot of $D_L(r)$ vs. $z(r)$.

Finally, noting that the factor in the denominator is the same quantity that appears in the density profile we can  rewrite $D_L$ as
\begin{eqnarray}
D_L=3\bb (1+z)\de rx_O^2\LT 1- {\e Lr_O\over 2\pi r \de r}\cos \LF 2\pi r\over L\RF\RT  \approx \nonumber \\
\approx 3\bb (1+z)\de rx_O^2\LT 1- {\e\over 2\pi}\LF{ L\over  \de r}\RF\cos \LF 2\pi r\over L\RF\RT \, ,
\end{eqnarray}
where the last approximation is only valid for the source close to the observer, but typically this is when the corrections are important. 
%%%%%%%%%%%%%%%%%%%%%%%%%%%%%%%%%%%%%%%
\section{Derivation using the Optical Scalar Equation}\label{sachs}

In this subsection we show that our result  ~(\ref{dlresult}) is consistent with the so-called optical scalar equations derived first by Sachs \cite{sachs}.
This equation has been used for example in the literature to study the distance $D_A$ when light travels through empty regions \cite{zeldovichL,kantowski,dyerroeder}.

The Sachs equation for the evolution of a cross sectional area $A$ for a cone of light originating from one point is
\be
\frac{1}{l}\frac{d^2l}{dv^2}=  {\cal R} \qquad \, \, {\rm where} \, \,\qquad {\cal R}\equiv \frac{1}{2}R_{\mu\nu}k^{\mu}k^{\nu} \,. \label{elle} \, .
\ee
Here $l\equiv \sqrt{A}$, $R_{\mu\nu}$ is the Ricci tensor and $k^{\mu}$ is the derivative of the position of the photon along its path:
$$k^{\mu}\equiv\frac{d x^{\mu}}{dv}=\left(\frac{dt(v)}{dv},\frac{dr(v)}{dv},0,0,\right).$$
We want to see if our result obeys this equation, in order to have an independent double-check.

We may express the derivatives in $k^{\mu}$ using again  ~(\ref{vt}) and  ~(\ref{t-radial}), so that:
\be
k^{\mu}=\left(-(1+z),\frac{\sqrt{1+2 E(r)}}{R'}(1+z),0,0\right) \, .
\ee

It is clear from the definitions that the quantity $l$ is proportional to the area distance $D_A$. Therefore $D_A$ has to obey the same equation~(\ref{elle}).

Now, it is straightforward to check that our solution  ~(\ref{dlresult}) obeys the following equation:
\be
\frac{1}{D_A}\frac{d^2 D_A}{dv^2}=\frac{1}{R}\frac{d^2 R}{dv^2} \, .
\ee
So we have to compare this with the term ${\cal R}$ of  ~(\ref{elle});
by direct computation the relevant components of the Ricci tensor read:
\begin{eqnarray}
R_{tt}&=& 2 \frac{\ddot{R}}{R}+ \frac{\ddot{R}'}{R'}\nonumber \, , \\
R_{rr}&=& \frac{1}{1+2 E(r)} \frac{R'(2 E'(r)-2 \dot{R}\dot{R}'-\ddot{R}' R)}{R} \, .
\end{eqnarray}
So, contracting with $k^{\mu}$ it is easy to see that in fact ${\cal R}=\frac{1}{R_A}\frac{d^2 R_A}{dv^2}$.

%%%%%%%%%%%%%%%%%%%%%%%%%%%%

\end{document}